\documentclass[a4paper,fleqn,usenatbib]{mnras}

\usepackage{newtxtext,newtxmath}

\usepackage[T1]{fontenc}
\usepackage{ae,aecompl}

\usepackage{graphicx}	
\usepackage{amsmath}	
\usepackage{amssymb}	

\title[PSR J2032]{The semicentennial binary system PSR J2032+4127 at periastron: X-ray photometry, optical spectroscopy and SPH modelling.}

\author[M.J.Coe et al]{M. J. Coe$^{1}$, A.T. Okazaki$^{2}$, I.A. Steele$^{3}$, C.-Y. Ng$^{4}$,
Wynn C.G. Ho$^{1,5,6}$, 
\newauthor A. G. Lyne$^{7}$, B. Stappers$^{7}$, T.J. Johnson$^{8}$, Paul S. Ray$^{9}$, M. Kerr$^{9}$.
\\
$^{1}$School of Physics and Astronomy, University of Southampton, Southampton SO17 1BJ, UK\\
$^{2}$Faculty of Engineering, Hokkai-Gakuen University, Toyohira-ku, Sapporo, 062-8605, Japan\\
$^{3}$Astrophysics Research Institute, Liverpool John Moores University, Liverpool, L3 5RF, UK\\
$^{4}$Department of Physics, the University of Hong Kong, Hong Kong, China\\
$^{5}$Department of Physics and Astronomy, Haverford College, 370 Lancaster Avenue, Haverford, PA 19041, USA\\
$^{6}$Mathematical Sciences and STAG Research Centre, University of Southampton, Southampton, SO17 1BJ, UK\\
$^{7}$ Jodrell Bank Centre for Astrophysics, School of Physics and Astronomy, University of Manchester, Manchester, M13 9PL, UK \\
$^{8}$ College of Science, George Mason University, Fairfax, VA 22030, USA  \\
$^{9}$ Space Science Division, U.S. Naval Research Laboratory, Washington, DC 20375-5352 USA\\
}

\date{Accepted XXX. Received YYY; in original form ZZZ}

\pubyear{2018}

\begin{document}
\label{firstpage}
\pagerange{\pageref{firstpage}--\pageref{lastpage}}
\maketitle

\begin{abstract}
X-ray photometry and optical spectra are presented covering the periastron passage of the highly-eccentric, $\sim$50 year binary system PSR J2032+4127 in November 2017. This system consists of a 143 ms pulsar in orbit around a massive OB star, MT 91-213. The data show dramatic changes during the encounter as the pulsar wind collided with the stellar wind. The X-ray flux rose on the approach to periastron, then underwent a major dip in the few days around periastron, and then gradually declined over the next few weeks. The optical spectroscopy revealed a steady decline in the H$\alpha$ line strength on the approach to periastron (from an Equivalent Width of -15{\AA} to -7{\AA}) implying a truncation of the OB star's circumstellar disk by the approaching neutron star. Smooth Particle Hydrodynamic (SPH) modelling is used here to model the system within the context of the observed behaviour and predict the geometrical configuration of the circumstellar disk with respect to the pulsar's orbit.
\end{abstract}




\section{Introduction}

The binary system PSR J2032+4127 is now well established as a 143 ms pulsar in partnership with the massive B0Ve star MT 91-213 \citep{mt91,camilo} in the Cygnus OB2 association.  The distance to MT91-213 is estimated at 1.38 kpc from GAIA data \citep{gaia}. The system has a binary period of $\sim$48 years and an eccentricity of 0.978 \citep{lyne2015, ho2017}. The pulsar was originally detected in a blind search through \textit{FERMI} $\gamma$-ray data \citep{abdo2009}, and subsequently the same period was detected in the radio \citep{camilo}. It is also associated with a \textit{MAGIC}, \textit{VERITAS} and\textit{HEGRA} TeV source \citep{aha2002, aha2005, aliu2014, abey2018}.The only other known similar system is PSR B1259-63, which hosts a 48 ms radio pulsar and an O9.5Ve star in a binary partnership with a period of 3.4 years (see, for example, \cite{dubus2013} for a review of the many papers written on this system).

The most recent periastron passage occurred on 13 November 2017 (=JD 2458070) and, given the long binary period, this offered the first opportunity to study this event with an array of telescopes at different wavelengths (see, for example, \citep{li2018, kolka2017, bednarek2018}). In this paper we present the X-ray measurements obtained by the \textit{Neil Gehrels Swift Observatory}  plus ground-based optical spectroscopy from the \textit{Liverpool Telescope}. Some of the X-ray data have already been presented in \cite{li2017, petro2018},  but the entire data set is combined here with new optical data and then used as the basis for Smooth Particle Hydrodynamical (SPH) simulations to help understand the system behaviour.

X-ray spectroscopic studies of the system using \textit{XMM-Newton} and \textit{NuStar} from the same multi-wavelength campaign are presented in a companion paper by \cite{ng2019}. In addition, detailed radio observations throughout the periastron passage may be found in \cite{lyne2019}.

\section{X-ray observations}

\begin{figure}
	\hspace{-0.3cm}
	\includegraphics[width=90mm,angle=0]{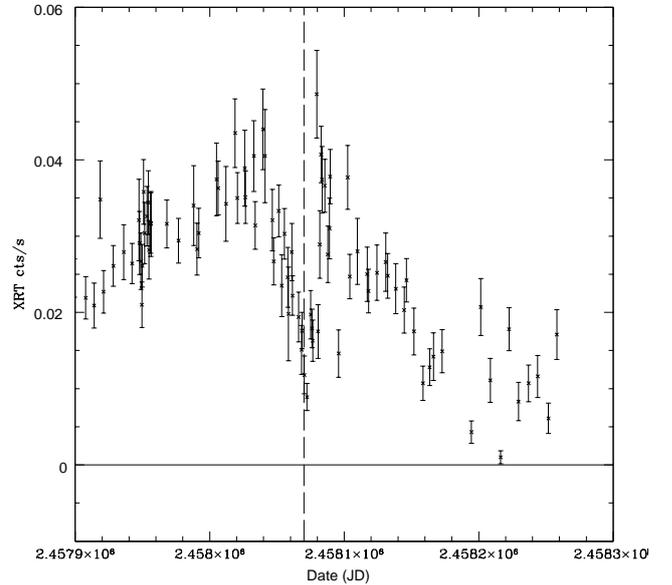}
    \caption{\textit{Neil Gehrels Swift Observatory} observatory detailed light curve around the time of periastron. The date of periastron (JD 2458070) is shown by the vertical dashed line.}
    \label{fig:xrt1}
\end{figure}

The source was monitored over the 0.3 - 10 keV range throughout the periastron passage by the  \textit{Neil Gehrels Swift observatory} \citep{gehrels2004}. The observations cover over a year and typical exposure times used were in the range 1-3 ks depending upon the spacecraft visibility window at that time. The XRT lightcurve was produced following the instructions described in the Swift data
analysis guide (http://www.swift.ac.uk/analysis/xrt/).  This resulted in collecting up to $\sim$100 background-subtracted counts per observation at the peak brightness just before the time of periastron, and then a flux very close to zero at the lowest post-periastron point - see Figure~\ref{fig:xrt1}. Though versions of the X-ray lightcurve have been published by \cite{li2018} and \cite{abey2018}, we reproduce the light curve again here as we wish to use it to make clear comparisons with our models later in this paper.

\begin{figure}
	\hspace{-0.3cm}
	\includegraphics[width=90mm,angle=0]{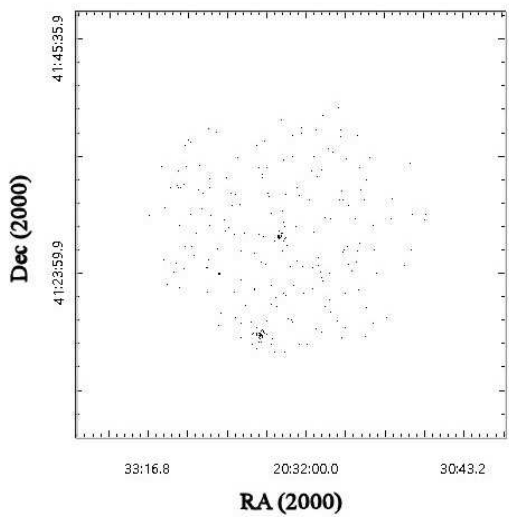}
    \caption{Typical \textit{Swift} image obtained on 13 Oct 2017 with an exposure time of 1580s.The full \textit{Swift} XRT field of view is shown and represents 23 x 23 arcminutes. PSR J2032+4127 is in the centre of the field and BD +40 4220 lies towards the southern edge.}
    \label{fig:xrt2}
\end{figure}

In addition to [MT 91] 213, the double supergiant binary system BD +40 4220 (=Schulte 5) was seen in almost every \textit{Neil Gehrels Swift Observatory} observation (except when it occasionally dropped off the edge of the field of view). See Figure~\ref{fig:xrt2} .

The X-ray measurements reveal a steady increase in the X-ray flux of PSR J2032+4127 on the approach to periastron, followed by a sharp dip around periastron. Then a subsequent sharp recovery and slow general decline after periastron. A possible second, even sharper dip is seen about 25d after periastron at JD 2458095.

\begin{figure}
	\hspace{-0.3cm}
	\includegraphics[width=80mm,angle=0]{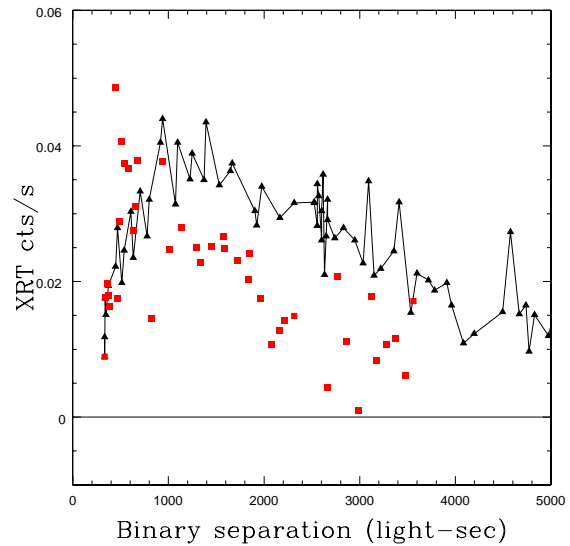}
    \caption{\textit{Neil Gehrels Swift Observatory} light curve versus binary separation. The black triangles show measurements pre-periastron, and the red squares show measurements obtained post-periastron. The size of the circumstellar disk around the OB star is $\sim$350 light-s. }
    \label{fig:sep}
\end{figure}

It is interesting to compare the X-ray flux before and after periastron in the context of the separation of the two objects - see Figure~\ref{fig:sep}. In this plot the X-ray measurements are shown as a function of the separation in light-seconds obtained using a Keplerian model of the orbit, and assuming an orbital inclination to the line-of sight of 60 degrees and a stellar mass ratio of 11.1 \citep{lyne2019}.  Globally the X-ray brightness on periastron approach is consistently much higher than post periastron for the same binary separation. The only exception to this pattern of behaviour is a short period immediately after periastron when the separation is 400-600 light-seconds.

\section{Optical Spectroscopy}

\begin{figure}
	\hspace{-0.3cm}
	\includegraphics[width=90mm,angle=0]{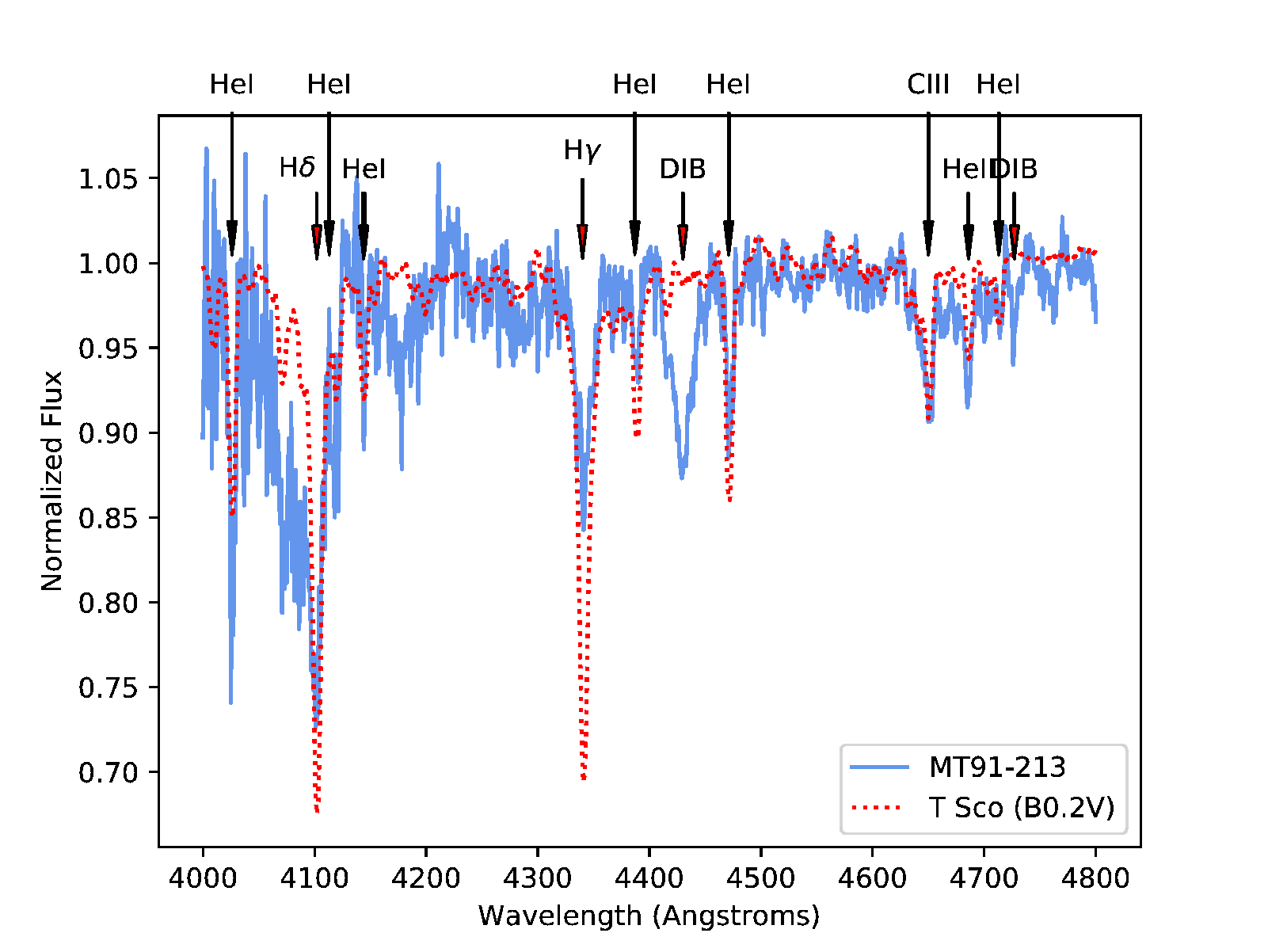}
    \caption{Classification Spectrum of MT91-213 (blue solid line) compared with the B0.2V spectral standard $\tau$ Sco (red dotted line).  The standard has been broadened to the same $v \sin i$ value as MT91-213 for ease of comparison.  The Balmer series lines were not used in classification due to infilling disk emission.  Regions affected by the diffuse interstellar bands (DIB) were also neglected.}
    \label{fig:blue-spec}
\end{figure}

\subsection{Observations}

Medium resolution optical spectroscopy of the counterpart of PSR J2032+4127 (MT91-213) was carried out using the fibre fed integral field FRODOSpec Spectrograph \citep{frodospec} on the 2.0m Liverpool Telescope \citep{lt} over the period 2016 April 25 -- 2018 April 28.  Observations were carried out using the high resolution VPH gratings in both the red and blue arms.  These provide wavelength coverage from 
3900--5100 and 5900-8000{\AA} 
with a resolution $R\sim5000$.   A total of 33 epochs of data were taken spread over the 2 year observing programme.  Data taking was intensified around the time of periastron, although poor weather prevented some observations at that time.  Total exposure time per epoch was 1800-s.  This was split into $3 \times 600$-s integrations to provide protection against cosmic rays.

Data reduction was carried out using the observatory supplied pipeline \citep{frodo-pipe}.  For each integration this involves the processes of bias removal, flat-fielding, distortion correction, fibre tracing, spectral extraction, sky subtraction and wavelength calibration.  No absolute flux calibration was carried out.  The three integrations per epoch were combined using median filtering to remove cosmic rays from the final spectra.  

\subsection{Spectral Type Reassessment}

The original classification of MT91-213 was made from a low resolution spectrum presented by \cite{mt91} in their Figure 11.  They obtained a spectral type of B0V and noted the line broadening due to rapid rotation.  Since our spectra are of higher resolution, we decided to re-visit this classification using the blue arm spectra.  Unfortunately the signal-to-noise ratio of the individual blue spectra is poor due to the low throughput of the instrument. So we therefore selected the four spectra corresponding to the time of weakest H$\alpha$ emission (Section \ref{sec:halpha}) and combined them together to make a 2 hour equivalent exposure that best represents the spectrum of the underlying B star.  This is presented in Figure \ref{fig:blue-spec} along with a spectrum of the B0.2V star $\tau$ Sco taken from the data set of \cite{be-stars}.  Comparisons were made with these and other spectra \citep{walborn}.  We note that the lack of He{\sc II} 4541 {\AA} rules out an earlier classification (e.g. O9.5V).  In our classification we did
not use the strength of the Balmer series lines due to contamination from circumstellar disk emission.  This visibly affects the H$\alpha$ and H$\beta$ lines and also contributes to some infilling of at least the H$\gamma$ line.  Overall we confirm that a spectral type in the range B0V -- B0.2V remains appropriate for MT91-213. 

We also used our spectrum to evaluate the rotation speed of the object using the the full-width half maximum (FWHM) -- rotation speed relationship of \cite{be-stars}. From the FWHM of the He{\sc i} 4026, 4387 and 4471 {\AA} lines we derived $v \sin i = 230 \pm 15 $ km/s, where the uncertainty reflects the spread in the values derived from the 3 lines.  If MT91-213 is rotating at the mass-shedding limit, then the escape velocity off the surface = 850 km/s. Thus we can estimate that $\sin i = 0.27 $ or  i $\ge$ $16^{\circ} $.

\subsection{Emission Lines}
\label{sec:halpha}

\begin{figure}
	\hspace{-0.3cm}
	\includegraphics[width=85mm,angle=0]{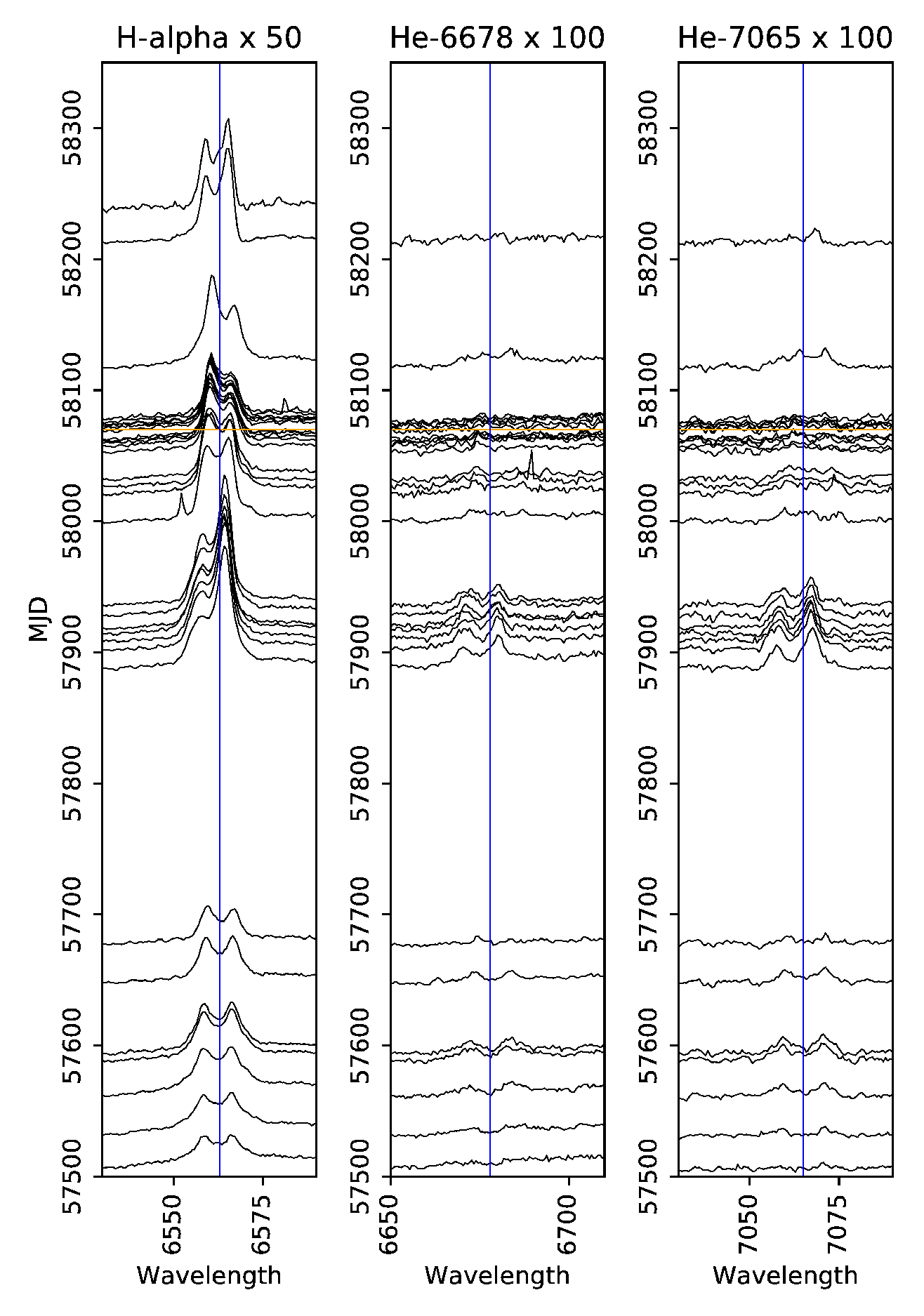}
    \caption{Line profiles of $H\alpha$, HeI 6678 {\AA} and HeI 7065{\AA}.  The spectra have been normalized to a continuum value of 1.0, offset by the observation MJD and then multiplied by the factor shown at the top of each column.  A small number of line profiles strongly affected by cosmic rays have been omitted.  The horizontal (orange) line indicates the date of periastron and the vertical (blue) line the rest wavelength in each case.}
    \label{fig:line-profiles}
\end{figure}

\begin{figure*}
	\hspace{-0.3cm}
	\includegraphics[width=150mm,angle=0]{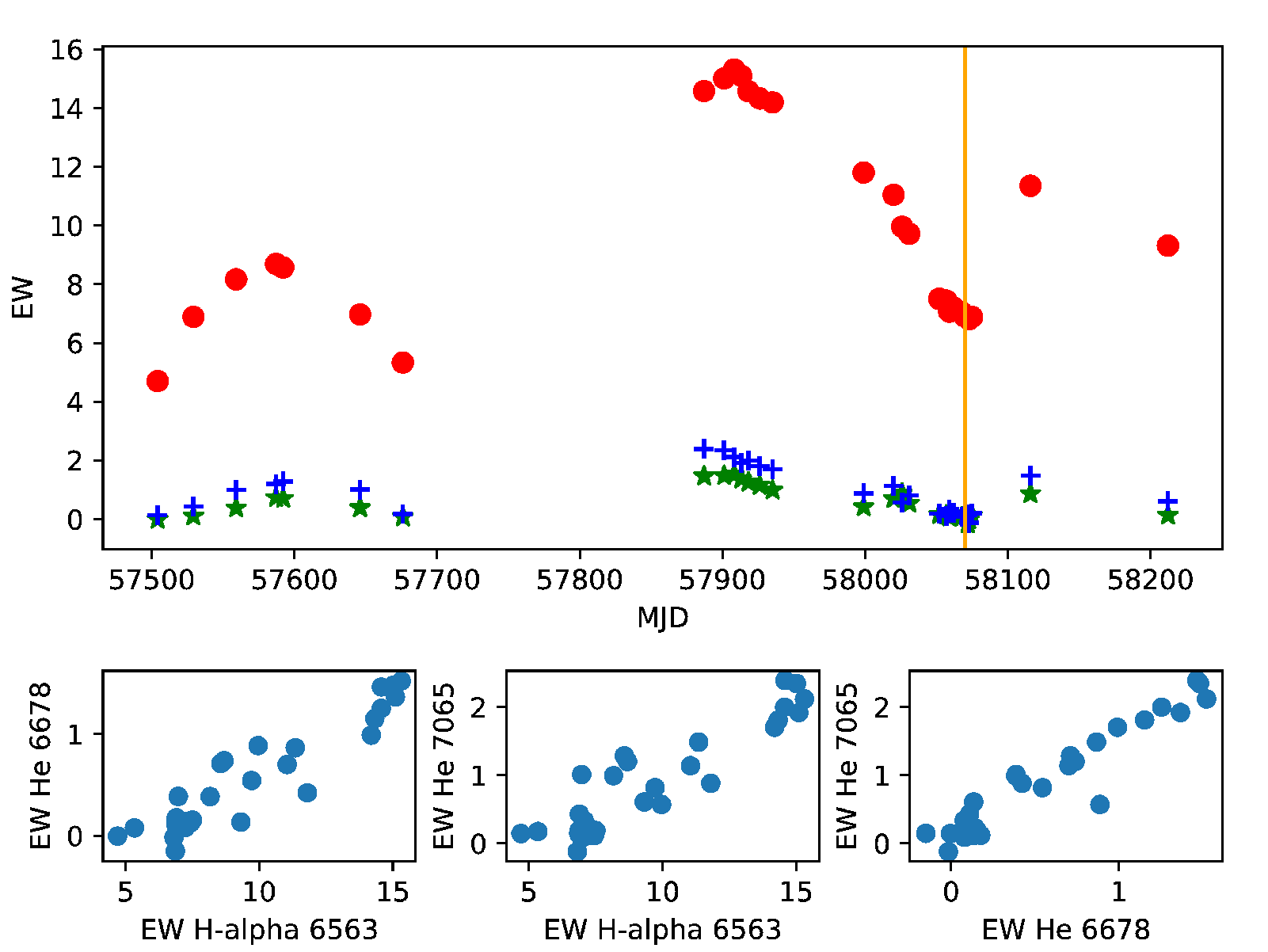}
    \caption{(top panel) Equivalent widths of H$\alpha$ (red solid circles), HeI 6678 {\AA} (green stars) and HeI 7065{\AA} (blue crosses) versus MJD.  The vertical (orange) line indicates the date of periastron. The lower three panels show the correlation of all three equivalent widths.}
    \label{fig:ew-comparison}
\end{figure*}

We confine our emission line analysis to the higher signal-to-noise, red-arm spectra.
In this region we see obvious emission from H$\alpha$, HeI 6678 {\AA} and HeI 7065 {\AA}.  Line profiles for all of these lines are presented in Figure \ref{fig:line-profiles}.  Visual inspection shows how the initially symmetrical double peaked line profiles of all three lines became strongly asymmetric in the year before periastron, before starting to return to a symmetrical state post periastron.  

Equivalent widths of all three lines were measured with respect to the nearby continuum.  In Figure \ref{fig:ew-comparison} we plot the equivalent widths of all three species as a function of time, and also against each other.  All three are well correlated (and so therefore likely formed in similar regions of the circumstellar disk).   The maximum equivalent width of all three species is around MJD 57910 and the minimum at MJD 58073 which is very shortly after periastron.  

As already described, it is clear that all of the line profiles are double peaked - as expected from a disk emission.  The H$\alpha$ profile also shows a very broad component at all times.  We found the H$\alpha$ profile was well fit by the sum of three Gaussian functions. We fitted all of the H$\alpha$ line profiles by such a combination, using least squares minimization to derive all fit parameters for a single epoch in parallel. Unfortunately the HeI series lines were too weak to fit robustly in most cases.   Our analysis therefore will concentrate only on H$\alpha$. The results of the fitting are presented in Table \ref{tab:haew}.  The three components consist of two narrow components and one broad component.  The narrow components (``blue'' and ``red'') have mean wavelengths of 6559.1 and 6566.0 {\AA}.  The mean line widths of the individual components are 1.6 (blue) and 1.4 {\AA} (red).  These standard deviations correspond to full width half maxima of 3.8 and 3.3 {\AA} respectively.

\begin{figure*}
	\hspace{-0.3cm}
	\includegraphics[width=150mm,angle=0]{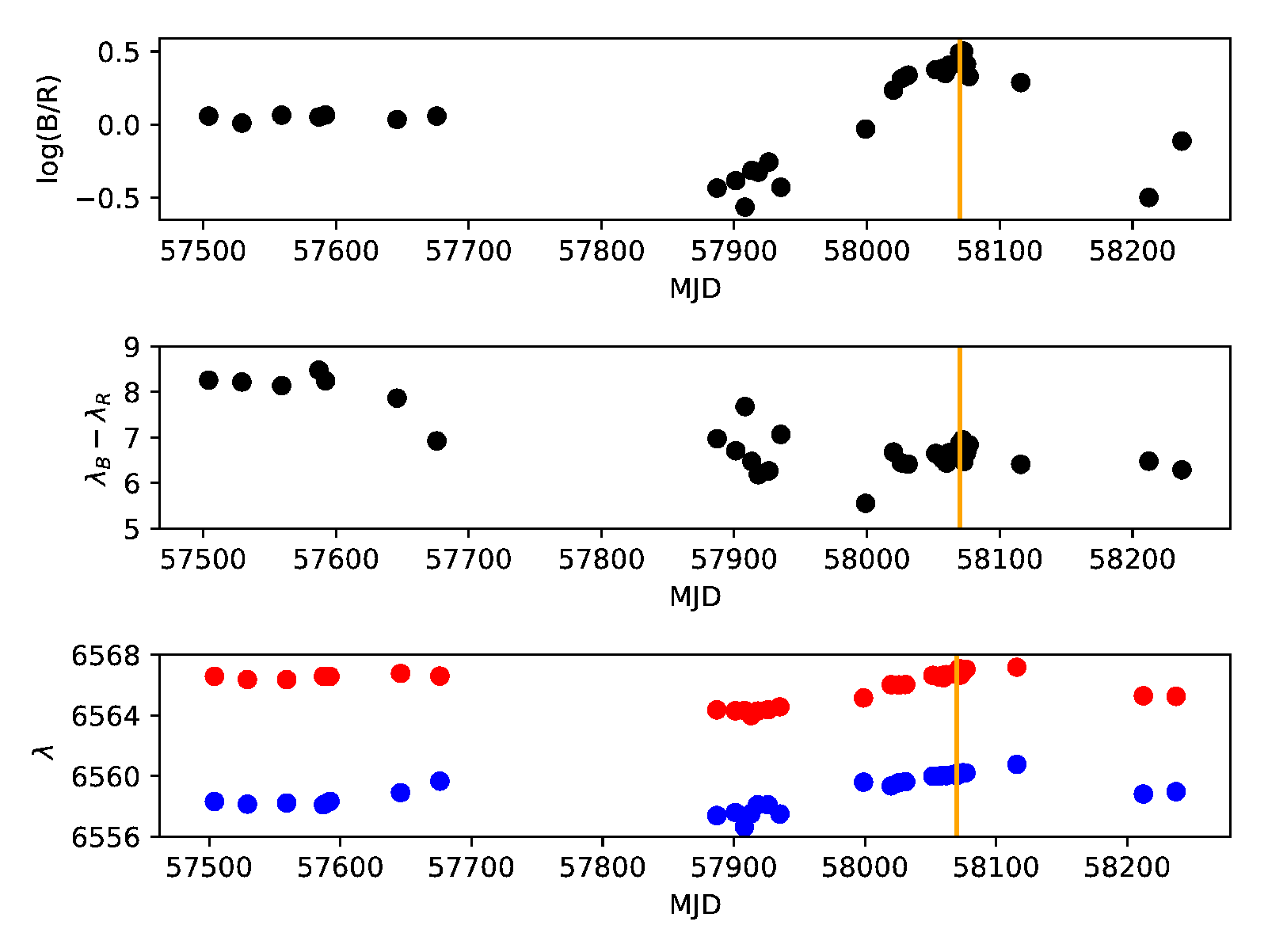}
    \caption{Properties derived from fits to the two narrow (``blue'' and ``red'') components of the H$\alpha$ line versus MJD.  The top panel shows the logarithm of the ratio of the peak flux of the components.  The centre panel shows the difference between the wavelengths of the two components. The lower panel shows the central wavelengths of each component. The vertical, orange line marks the time of periastron.  }
    \label{fig:combined}
\end{figure*}

In the upper panel of Figure \ref{fig:combined} we plot the ratio of the peak height of the fitted blue and red components, whereas this quantity was stable in 2016, in 2017-18 significant B/R variability is apparent.  The rise to a maximum of this ratio at the time of periastron is striking indicating significant circumstellar disk distortion.

The mean separation between the two narrow components is  6.9 {\AA} which corresponds to a velocity of 315 km/s.  Over the course of the entire data set this quantity decreases from a mean of $\sim 8$ {\AA} in 2016 to $\sim6$ {\AA} in 2018 (Figure \ref{fig:combined} - centre panel).

The wide component has mean wavelength 6563.2 {\AA} and a mean full width half maximum of 12.6 {\AA}.  This width is dominated by electron scattering and should not be interpreted kinematically \citep{han89}.

In Figure \ref{fig:combined} (lower panel) we plot how the mean wavelength of the two narrow Gaussian components varies with time.  In comparison with the 2016 data (MJD<57700) the 2017 data shows considerable variation of these quantities.  At the start of the observing season (MJD$\sim57900$) the blue and red components are observed at wavelength considerably bluer than their 2016 mean.  These components both then move towards the red end of the spectrum over the course of the year until by periastron they reach a maximum.  By 2018 the values appear to have returned closer to their 2016 mean.  In contrast the wavelength of the wide component remains stable over the whole time period.

\subsection{Circumstellar disk size}

For a Keplerian disk a simple estimate of the radius of the circumstellar disk ${R_{CD}}$ may be made. The peak separation of the H$\alpha$ component is 315 km/s which, if we interpret that as a Doppler shift, gives a Keplerian disk velocity of $\sim 160$ km/s. Assuming $\sin i = 1$ then this corresponds to a disk radius of $R_{CD}\sim15R_{OB} \sim 0.7 AU $.

Assuming an orbital period of 48 years and an eccentricity of 0.978 then the periastron distance will be 0.73 AU. So the pulsar's closest approach to the OB star is close to the edge of the circumstellar disk revealed by the H$\alpha$ measurements. Plus the occurrence during 2017 of a strong $B/R$ peak cycle, the equivalent width minimum at periastron and the correlated wavelength shift of the two components over 2017 all point to the influence of the neutron star passage on the disk \citep{reig}, with tidal forces initiating a disk asymmetry \citep{han94,o97}.

\begin{table*}
\label{tab:haew}
\caption{Journal of parameters derived from the H$\alpha$ line.  EW$_{\rm raw}$ is a standard equivalent width measurement derived from the raw data without fitting.  The subsequent columns list the properties of three Gaussian fits corresponding to the blue (1) and red (2) narrow components and the wide (3) component.  The columns list central wavelength ($\lambda$, {\AA}), standard deviation ($\sigma$, {\AA}), peak height (A) and integrated area under the fits, corresponding to -EW, ({\AA}) for each of the three components.  All properties are measured with respect to a normalized continuum.}
\begin{tabular}{lrrrrrrrrrrrrr}
\hline
MJD &  EW$_{\rm raw}$ & $\lambda_1$ & $\lambda_2$ & $\lambda_3$ & $\sigma_1$ & $\sigma_2$ & $\sigma_3$ & $A_1$ &  $A_2$ &  $A_3$ &  -EW$_1$ &   -EW$_2$ & -EW$_3$ \\
\hline
57504.115590 &  -4.70 &  6558.32 &  6566.58 &  6563.05 &  1.49 &  1.24 &   7.44 &  0.16 &  0.14 &  0.23 &  0.58 &  0.42 &   4.21 \\
57529.199445 &  -6.89 &  6558.14 &  6566.36 &  6562.97 &  1.49 &  1.21 &   7.65 &  0.18 &  0.18 &  0.34 &  0.68 &  0.54 &   6.44 \\
57559.154624 &  -8.17 &  6558.22 &  6566.35 &  6563.05 &  1.63 &  1.39 &   7.32 &  0.24 &  0.21 &  0.39 &  0.99 &  0.73 &   7.25 \\
57587.112859 &  -8.69 &  6558.10 &  6566.58 &  6563.44 &  1.73 &  1.46 &   6.60 &  0.32 &  0.29 &  0.41 &  1.40 &  1.05 &   6.84 \\
57592.130870 &  -8.57 &  6558.32 &  6566.57 &  6563.54 &  1.79 &  1.56 &   6.69 &  0.35 &  0.30 &  0.39 &  1.56 &  1.17 &   6.47 \\
57646.018603 &  -6.97 &  6558.90 &  6566.77 &  6563.41 &  1.55 &  1.41 &   6.02 &  0.33 &  0.30 &  0.33 &  1.27 &  1.07 &   5.03 \\
57675.963940 &  -5.33 &  6559.67 &  6566.59 &  6566.67 &  2.12 &  1.97 &  12.81 &  0.43 &  0.38 &  0.06 &  2.29 &  1.85 &   1.99 \\
57887.118891 & -14.58 &  6557.39 &  6564.36 &  6563.50 &  2.80 &  1.79 &   6.92 &  0.48 &  1.31 &  0.36 &  3.39 &  5.87 &   6.18 \\
57901.179391 & -15.01 &  6557.59 &  6564.30 &  6563.26 &  2.72 &  1.74 &   6.90 &  0.61 &  1.48 &  0.30 &  4.17 &  6.42 &   5.19 \\
57908.149722 & -15.32 &  6556.64 &  6564.32 &  6561.72 &  1.82 &  1.47 &   5.24 &  0.30 &  1.10 &  0.79 &  1.37 &  4.05 &  10.40 \\
57913.168435 & -15.10 &  6557.52 &  6563.99 &  6562.65 &  2.64 &  1.89 &   7.86 &  0.64 &  1.32 &  0.28 &  4.24 &  6.24 &   5.60 \\
57918.187500 & -14.57 &  6558.10 &  6564.28 &  6562.97 &  2.73 &  1.70 &   6.96 &  0.66 &  1.41 &  0.27 &  4.55 &  5.97 &   4.77 \\
57926.054738 & -14.33 &  6558.10 &  6564.37 &  6563.72 &  2.84 &  1.75 &   6.49 &  0.70 &  1.27 &  0.27 &  5.01 &  5.54 &   4.42 \\
57935.133659 & -14.19 &  6557.50 &  6564.56 &  6561.74 &  1.62 &  1.38 &   4.66 &  0.43 &  1.16 &  0.75 &  1.75 &  4.01 &   8.77 \\
57998.943637 & -11.81 &  6559.59 &  6565.15 &  6562.67 &  1.97 &  1.76 &   7.05 &  0.85 &  0.91 &  0.23 &  4.21 &  4.01 &   4.06 \\
58019.913414 & -11.05 &  6559.34 &  6566.02 &  6563.46 &  1.43 &  0.81 &   4.19 &  0.56 &  0.33 &  0.81 &  2.01 &  0.66 &   8.52 \\
58025.919956 &  -9.96 &  6559.56 &  6566.00 &  6563.94 &  1.68 &  0.99 &   4.01 &  0.59 &  0.29 &  0.69 &  2.50 &  0.71 &   6.95 \\
58030.955914 &  -9.72 &  6559.62 &  6566.04 &  6563.94 &  1.55 &  0.97 &   4.08 &  0.56 &  0.26 &  0.69 &  2.18 &  0.62 &   7.07 \\
58051.819740 &  -7.50 &  6559.99 &  6566.64 &  6563.53 &  1.34 &  1.30 &   3.88 &  0.53 &  0.22 &  0.53 &  1.77 &  0.73 &   5.13 \\
58056.825851 &  -7.45 &  6560.00 &  6566.53 &  6563.48 &  1.33 &  1.27 &   3.95 &  0.48 &  0.20 &  0.54 &  1.59 &  0.63 &   5.33 \\
58058.916801 &  -7.07 &  6560.03 &  6566.64 &  6563.35 &  1.37 &  1.49 &   3.71 &  0.52 &  0.23 &  0.49 &  1.77 &  0.86 &   4.52 \\
58059.935327 &  -7.24 &  6560.05 &  6566.48 &  6563.48 &  1.30 &  1.24 &   4.01 &  0.54 &  0.23 &  0.48 &  1.75 &  0.72 &   4.88 \\
58061.835944 &  -7.21 &  6560.03 &  6566.69 &  6563.34 &  1.23 &  1.28 &   3.86 &  0.50 &  0.20 &  0.53 &  1.55 &  0.63 &   5.10 \\
58067.823851 &  -7.04 &  6560.11 &  6566.75 &  6563.31 &  1.27 &  1.35 &   3.86 &  0.53 &  0.20 &  0.49 &  1.70 &  0.69 &   4.72 \\
58069.811054 &  -6.90 &  6560.03 &  6566.90 &  6563.35 &  1.17 &  1.11 &   3.80 &  0.51 &  0.17 &  0.53 &  1.50 &  0.46 &   5.05 \\
58071.832268 &  -6.87 &  6560.13 &  6567.08 &  6563.43 &  1.27 &  1.12 &   3.94 &  0.48 &  0.17 &  0.50 &  1.51 &  0.47 &   4.96 \\
58072.898812 &  -6.81 &  6560.19 &  6566.66 &  6563.46 &  1.22 &  1.15 &   3.80 &  0.52 &  0.16 &  0.51 &  1.59 &  0.47 &   4.85 \\
58074.870135 &  -6.89 &  6560.23 &  6566.89 &  6563.44 &  1.20 &  1.24 &   3.83 &  0.47 &  0.18 &  0.52 &  1.41 &  0.56 &   5.02 \\
58076.930428 &  -6.22 &  6560.21 &  6567.05 &  6563.33 &  1.13 &  1.12 &   3.50 &  0.42 &  0.20 &  0.53 &  1.18 &  0.55 &   4.68 \\
58115.828472 & -11.35 &  6560.76 &  6567.17 &  6563.07 &  1.31 &  1.27 &   5.62 &  0.70 &  0.36 &  0.60 &  2.29 &  1.14 &   8.43 \\
58212.246712 &  -9.32 &  6558.82 &  6565.29 &  6560.57 &  0.79 &  1.31 &   3.16 &  0.35 &  1.10 &  0.67 &  0.69 &  3.60 &   5.32 \\
58237.167561 &  -7.84 &  6558.97 &  6565.25 &  6562.51 &  1.50 &  1.19 &   1.01 &  0.89 &  1.15 &  0.59 &  3.36 &  3.45 &   1.50 \\
\hline
\end{tabular}

\end{table*}

\section{SPH Simulations}
\label{sec:SPH}

\citet{Takata2012} showed that the temporal change in the X-ray emission from PSR~B1259$-$63/LS~2883 around periastron is reproduced by a model where the hydrodynamic simulations of the pulsar wind interacting with the Be-star disk and wind are used to calculate the synchrotron emission from the shocked pulsar wind. It is interesting to study whether the same approach works for PSR~J2032+4127/MT 91~213 and, if it works, to probe Be-disk parameters, such as the orientation, the density distribution, and the outer radius.

\subsection{Numerical Setup}
\label{sec:hydro-model}

The simulations are performed with the same SPH code as in \citet{Okazaki2011} and \citet{Takata2012}. 
It was developed originally by \citet{Bate1995} and then modified so as to work for Be decretion disks and stellar winds \citep{Okazaki2002, Okazaki2008, Okazaki2011}. 
In the code, the Be circumstellar disk, the Be-star wind, and the pulsar wind are modeled by ensembles of gas particles of different particle masses with negligible self-gravity, whereas the Be star and the pulsar are represented by sink particles with the appropriate gravitational masses. 
Gas particles which fall within a specified accretion radius are accreted by the sink particle.
The radiative cooling is taken into account with the cooling function generated by CLOUDY 90.01 for an optically thin plasma with solar abundances \citep{Ferland1996}.
We set the binary orbit as the $x$-$y$ plane with the apastron and periastron in the $+x$- and $-x$-directions, respectively.

\begin{figure}
	\resizebox{\hsize}{!}{\includegraphics{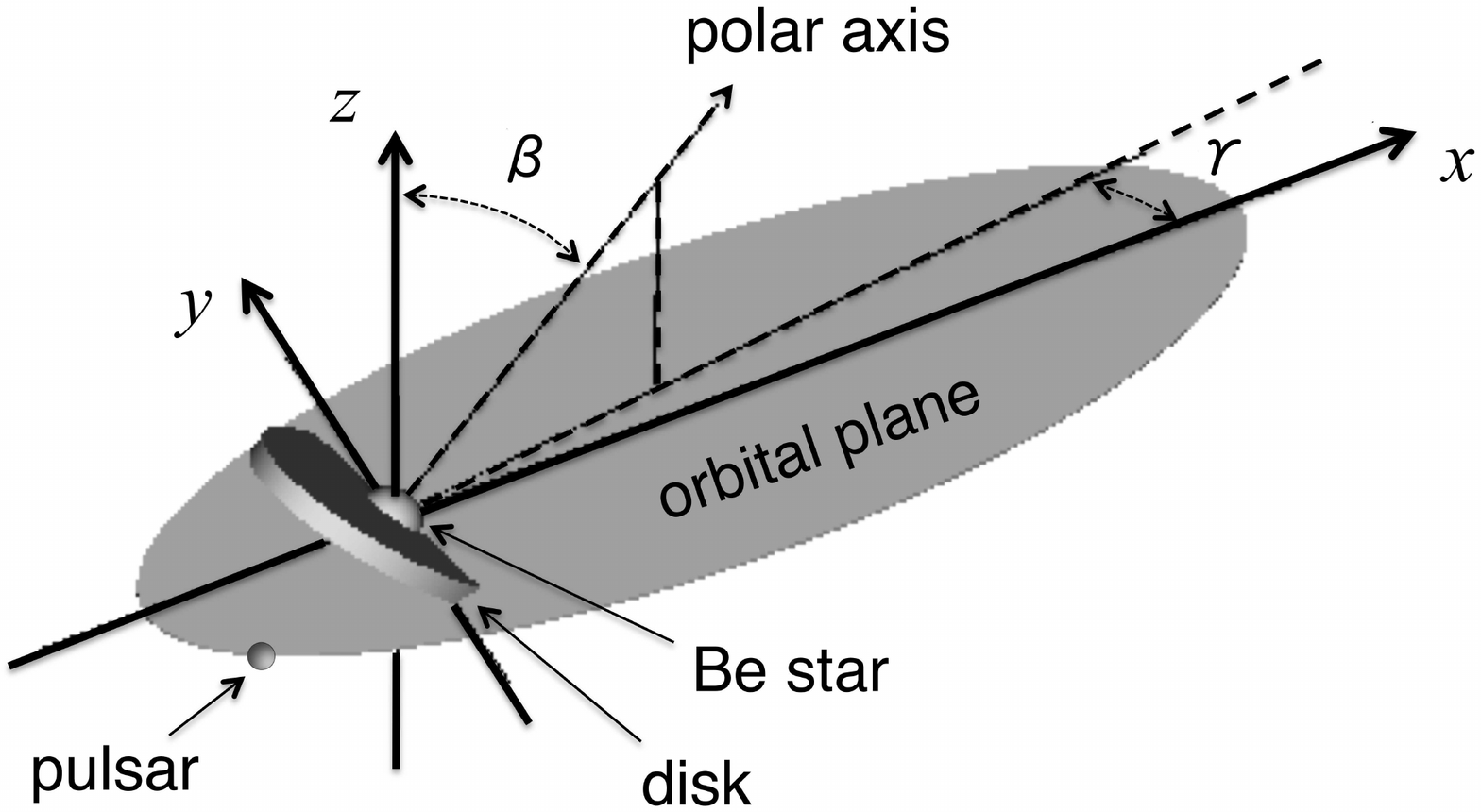}}
    \caption{Model geometry of PSR J2032$+$4127/MT91~213.
    The binary orbit is in the $x$-$y$ plane. The polar axis of the Be star is tilted 
    by $\beta$ to the $z$-axis and has the azimuth of tilt, $\gamma$, to
    the $x$-axis. The diagram is not to scale.}
    \label{fig:geometry}
\end{figure}

In each simulation, we first set up a Be disk around a spherical Be star, which is inclined to the binary orbit plane with the tilt angle, $\beta$, and the azimuth of tilt measured from the $x$-axis (i.e., the apastron direction), $\gamma$ (see Fig.~\ref{fig:geometry} for model geometry). The initial disk has a typical density distribution in the form, $\rho_0 (r/R_\mathrm{OB})^{-3.5}$, in the equatorial plane of the Be star and is in hydrostatic equilibrium in the direction perpendicular to it. Here, $\rho_0$ is the quantity called the base density and $R_\mathrm{OB}$ is the radius of the Be star.
The disk temperature is assumed to be isothermal at the temperature of $0.6\,T_\mathrm{eff}$ \citep{Carciofi2006}. The initial disk consists of $10^5$ particles.

Then, we start a simulation at $t = -0.01 P_\mathrm{orb}$ ($\sim 175$ days prior to periastron). 
The mass ejection from the star, the Be wind, and the pulsar wind are turned on at that moment.
We emulate the mass ejection from the Be star by injecting particles with the local Keplerian speed at a constant rate of $10^{-8} M_\odot\,\mathrm{yr}^{-1}$ at a radius just outside the Be star's equator.
For simplicity, we assume that the winds coast without any net external force, 
assuming in effect that gravitational forces are 
either negligible (i.e. for the pulsar wind) or are cancelled by radiative driving terms (i.e. for the Be wind).
The relativistic pulsar wind is emulated by a non-relativistic flow with a velocity of $10^{4}\,{\rm km\;s}^{-1}$ and an adjusted mass-loss rate so as to provide the same momentum flux as a relativistic flow with the same assumed energy.
We assume that all the spin down energy 
$L_{\rm sd}=1.5 \times 10^{35}\,{\rm erg~s}^{-1}$ goes into 
the kinetic energy of a spherically symmetric pulsar wind. 
We also assume the Be wind to be spherically symmetric with 
the mass loss rate $\dot{M}$ and a constant velocity $V_\mathrm{wind}$, 
both of which are parameters given in each simulation.
We run simulations until $t = 0.01 P_\mathrm{orb}$ ($\sim 175$ days after periastron).

We study three models, of which the parameters are summarized in Table~\ref{tab:models}. These models have different sets of the initial azimuthal direction, $\gamma$, and radius, $r_\mathrm{disk}$, of the Be disk and the Be-wind velocity, $V_\mathrm{wind}$: In Models~1 and 2, the initial disk radius is the same, but the disk azimuth and wind velocity are different. In Models~2 and 3, on the other hand, only the initial disk radius is different. In all models, the other Be-star parameters are fixed, as well as those of the pulsar.

When the global form of the pulsar wind interacting with the stellar wind is considered, a key parameter is the ratio of wind momentum fluxes, $\eta$, given by
\begin{eqnarray}
\eta &=& \frac{L_\mathrm{sd}}{\dot{M} V_\mathrm{wind}\,c} \nonumber\\
     &=& 0.053 \left( \frac{L_\mathrm{sd}}{10^{35}\,{\rm erg~s}^{-1}} \right) \nonumber\\
     && \times \left( \frac{\dot{M}}{10^{-8}\,M_\odot\,\mathrm{yr}^{-1}} \right)^{-1}
       \left( \frac{V_\mathrm{wind}}{10^3\,\mathrm{km\;s}^{-1}} \right)^{-1}.
\label{eq:eta}
\end{eqnarray}
Using the parameters adopted above, we have $\eta = 0.08$ in Model~1 and $\eta = 0.04$ in Models~2 and 3. Consequently, the collisional shock between the pulsar and stellar winds is wrapped around the pulsar in all the three models studied below.

\begin{table}
\caption{Model parameters. The inclination of the binary orbit plane to the circumstellar disk plane is characterized by the angle, $\beta$, and the azimuth of tilt measured from the $x$-axis (i.e., the apastron direction), is $\gamma$. The parameter $r_\mathrm{disk}/a$ is the initial disk radius normalized by the periastron distance.}
\begin{tabular}{@{}ccccc@{}ccc}
\hline
Model & \multicolumn{4}{c}{Be disk} && \multicolumn{2}{c}{Stellar wind} \\
\cline{2-5} \cline{7-8}
& $\beta$ & $\gamma$ & $r_\mathrm{disk}/a$ & $\rho_{0}$ &&
$\dot{M}$ & $V_\mathrm{wind}$ \\
&&&& (g/cm$^{3}$) && ($M_\odot/\mathrm{yr}$) & (km/s) \\
\hline
Model 1 & $45^\circ$ & $-45^\circ$ & 0.02 & $10^{-11}$ && $10^{-8}$ & $10^3$ \\
Model 2 & $45^\circ$ & $+30^\circ$ & 0.02 & $10^{-11}$ && $10^{-8}$ & $2 \cdot 10^3$ \\
Model 3 & $45^\circ$ & $+30^\circ$ & 0.1 & $10^{-11}$ && $10^{-8}$ & $2 \cdot 10^3$ \\
\hline
\end{tabular}
\label{tab:models}
\end{table}

\subsection{Hydrodynamic Interaction between the Pulsar Wind and the Disk and Wind of the Be Star}
\label{sec:simulations}

\begin{figure*}
	\vspace{0.5cm}
	\resizebox{\hsize}{!}{
	\includegraphics[height=38mm,angle=0]{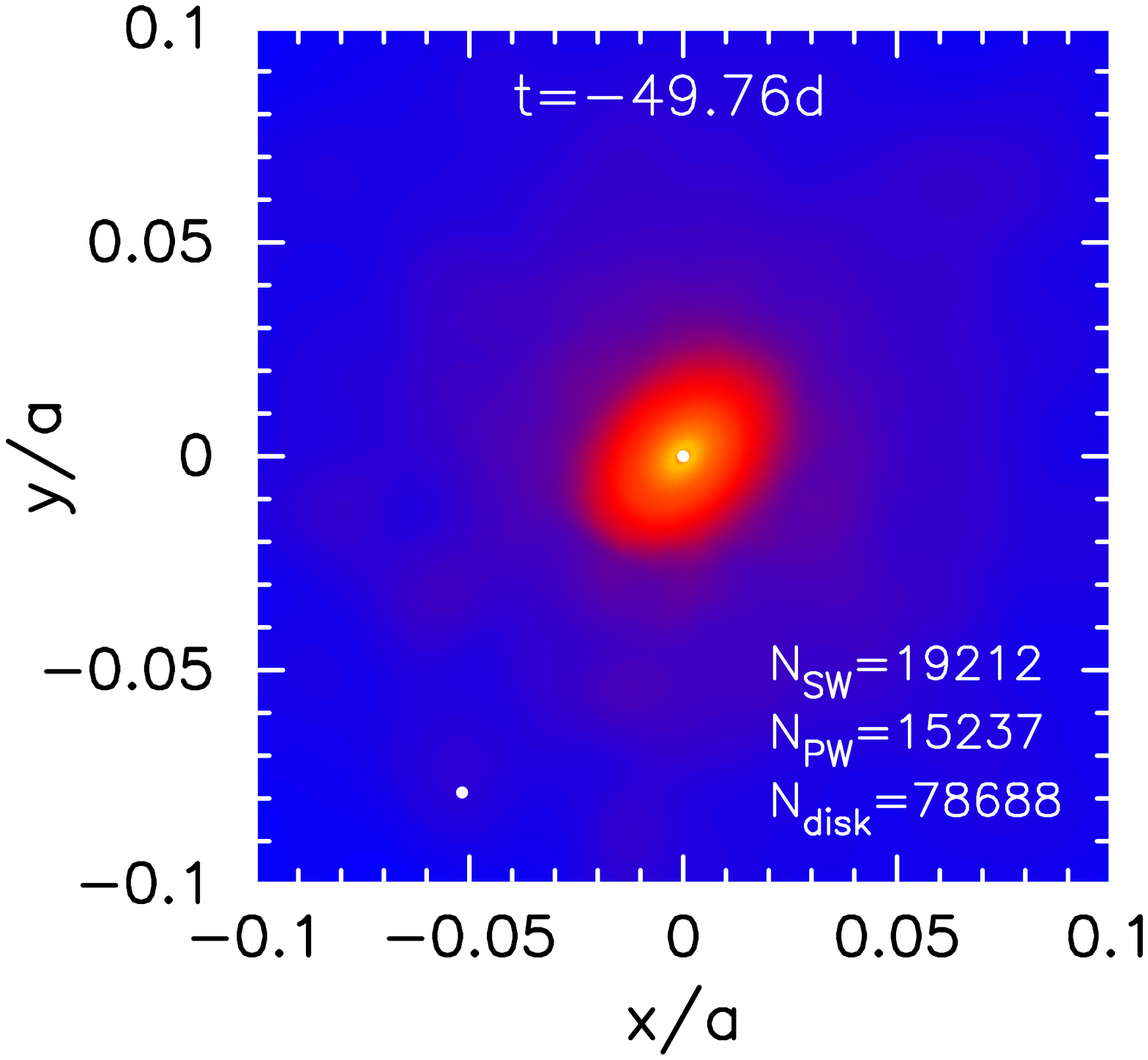} 
	\includegraphics[height=38mm,angle=0]{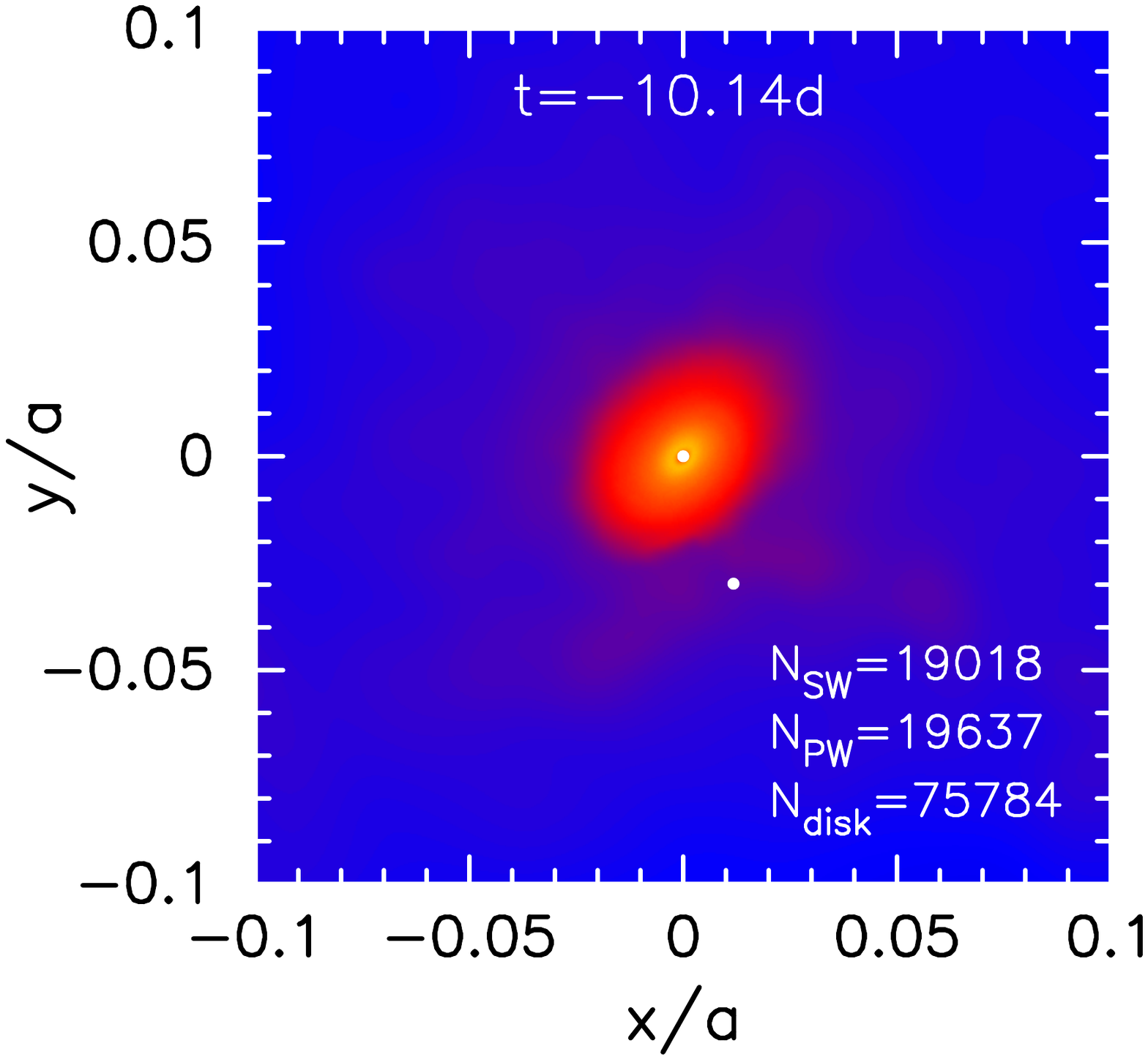} 
	\includegraphics[height=38mm,angle=0]{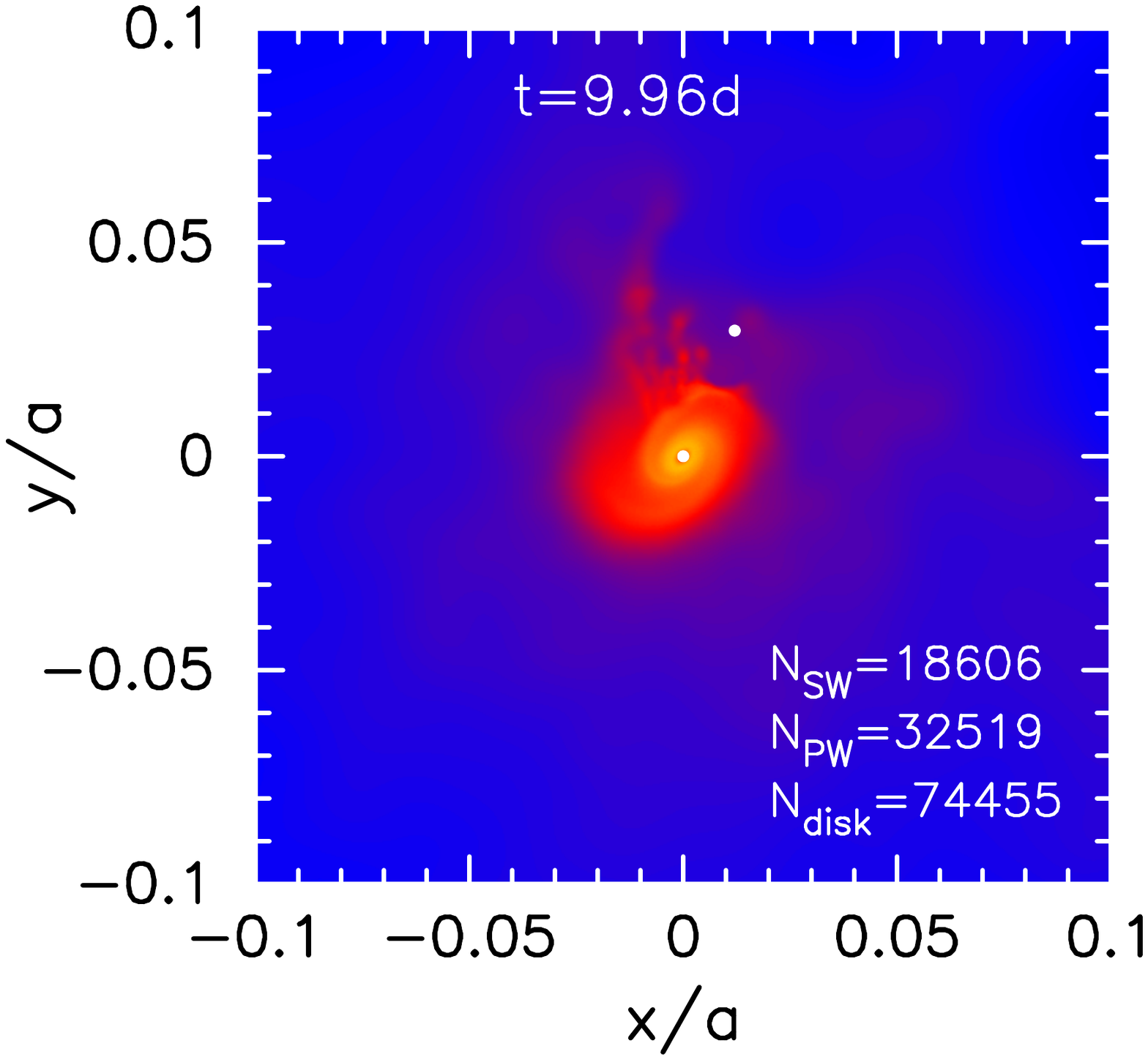} 
	\includegraphics[height=38mm,angle=0]{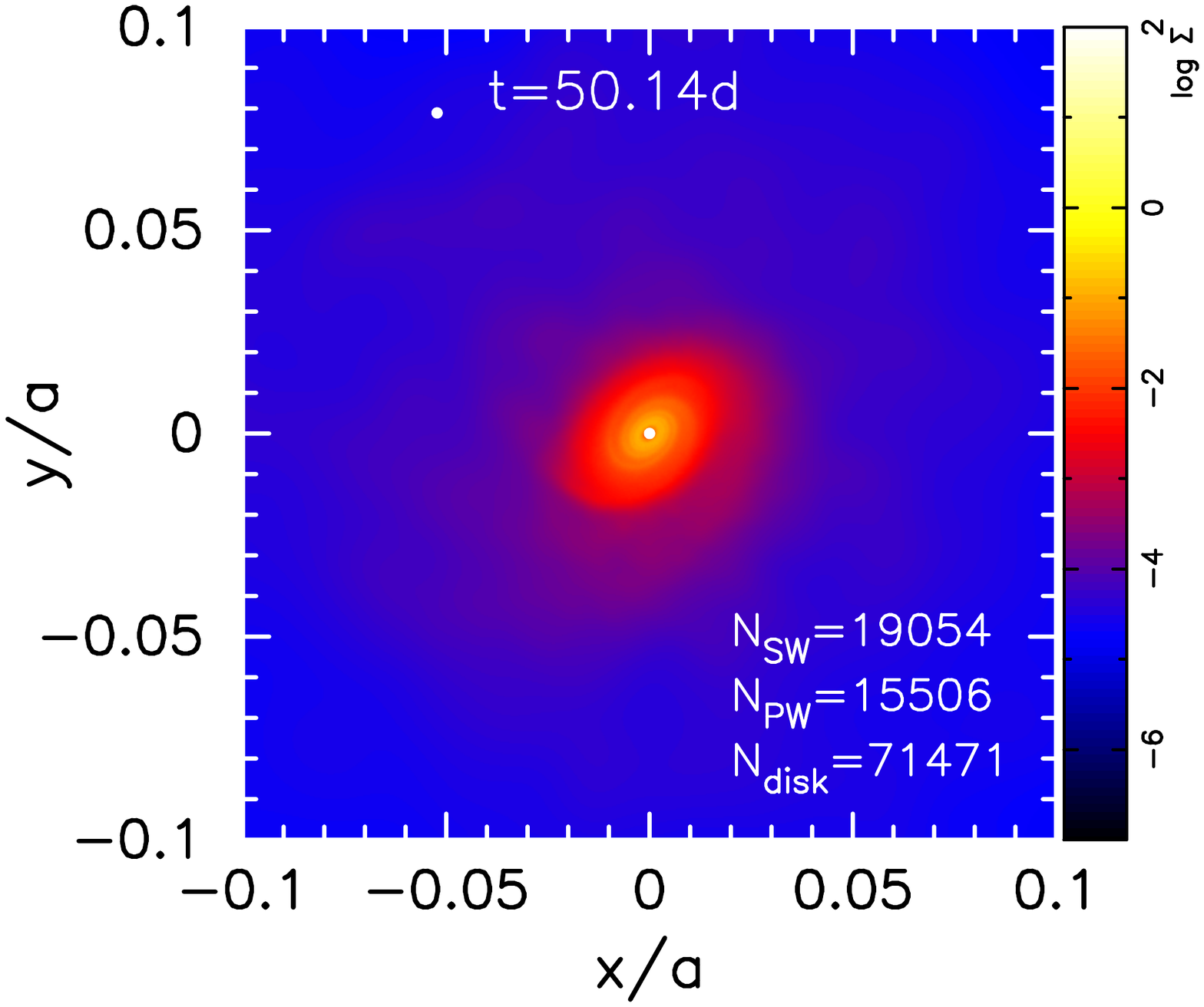}}
	\resizebox{\hsize}{!}{
	\includegraphics[height=38mm,angle=0]{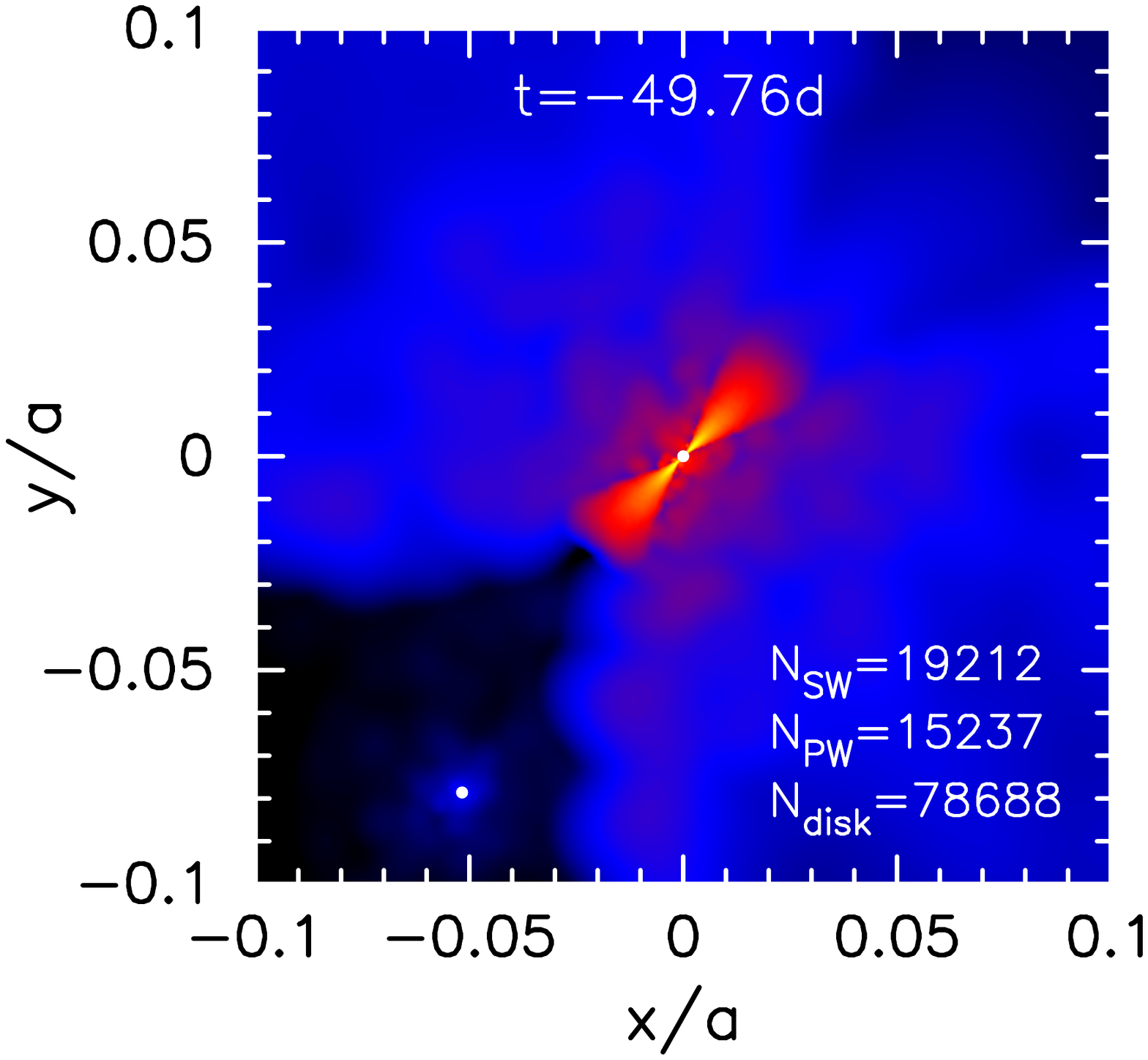} 
	\includegraphics[height=38mm,angle=0]{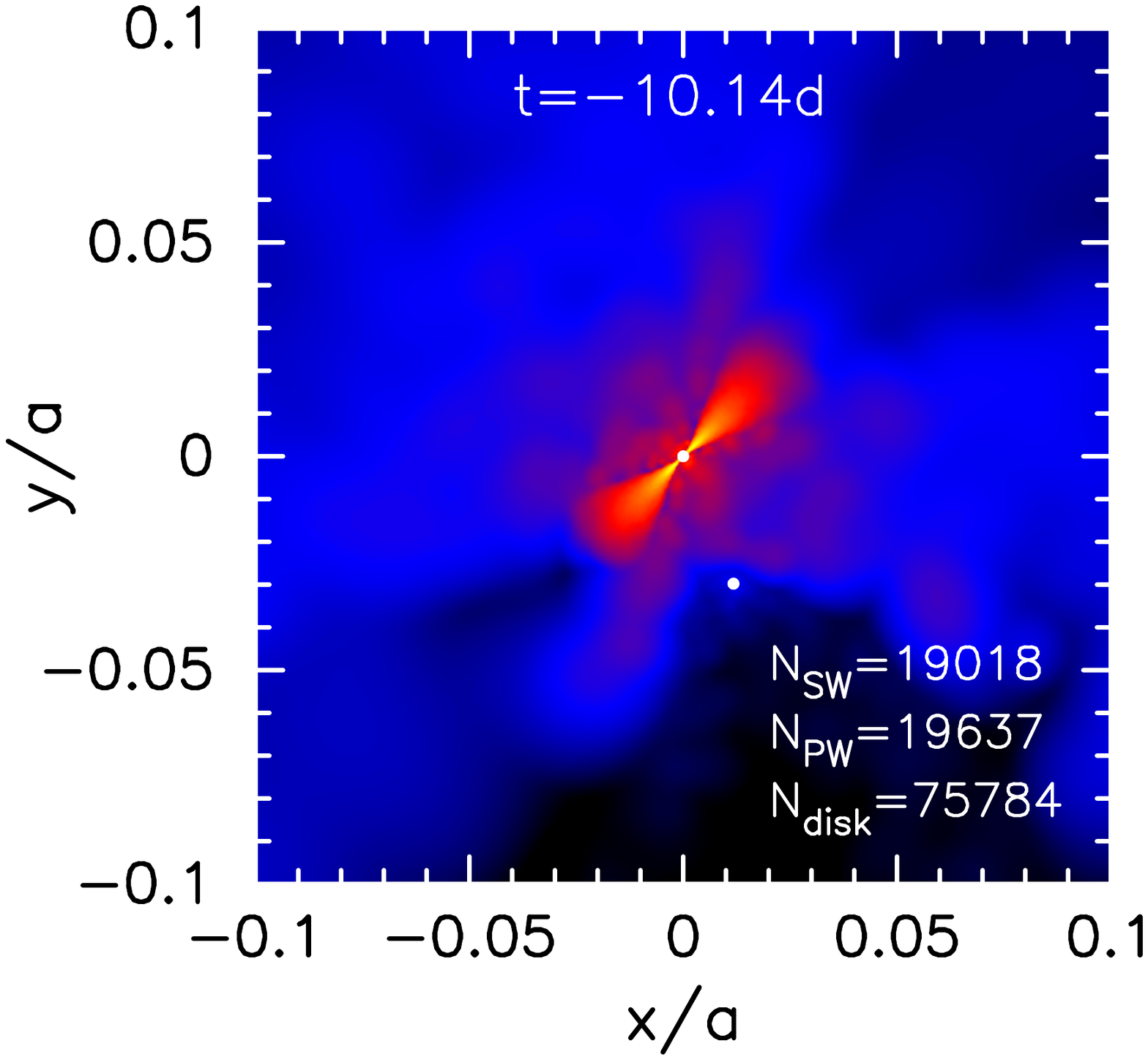} 
	\includegraphics[height=38mm,angle=0]{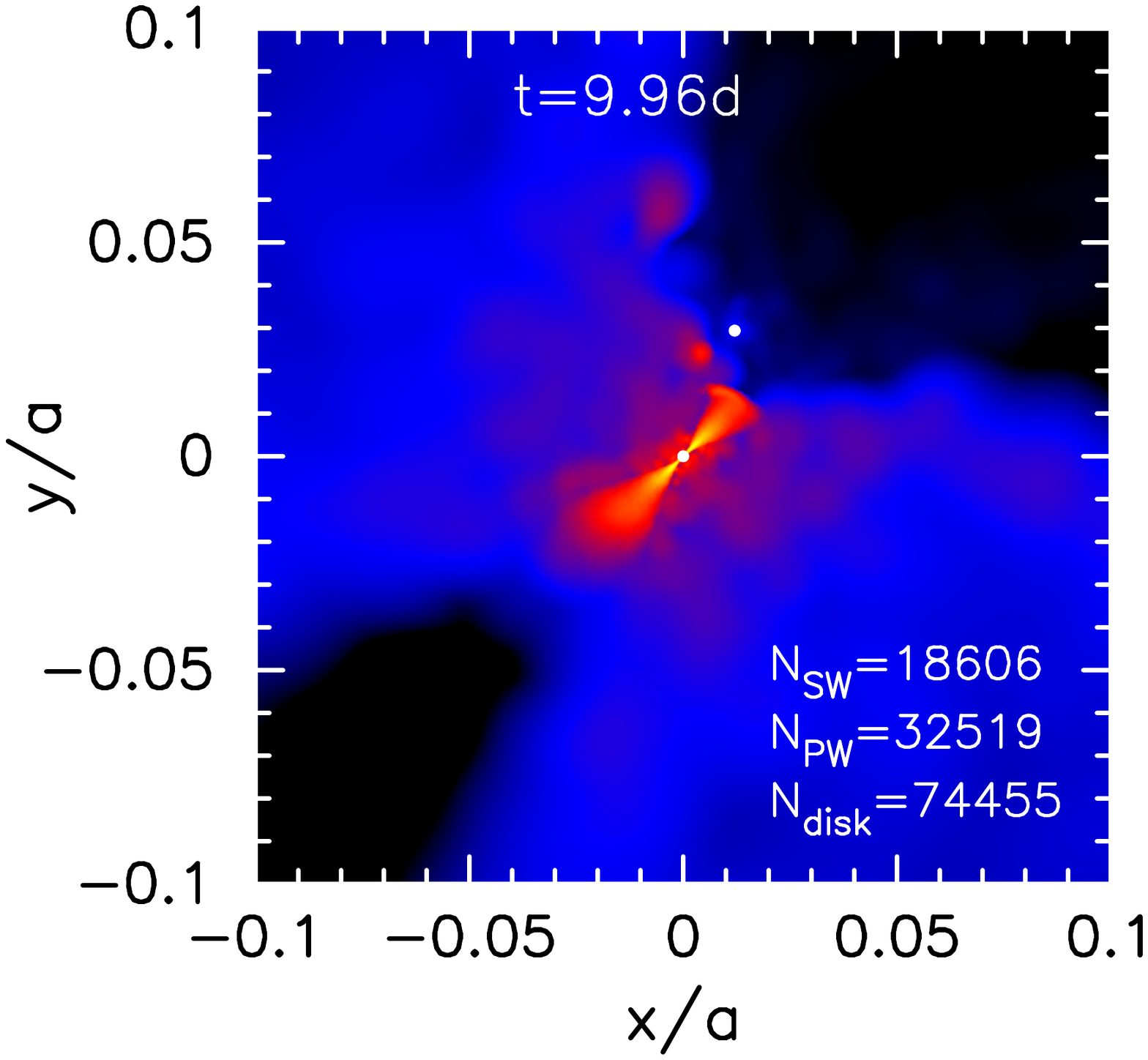} 
	\includegraphics[height=38mm,angle=0]{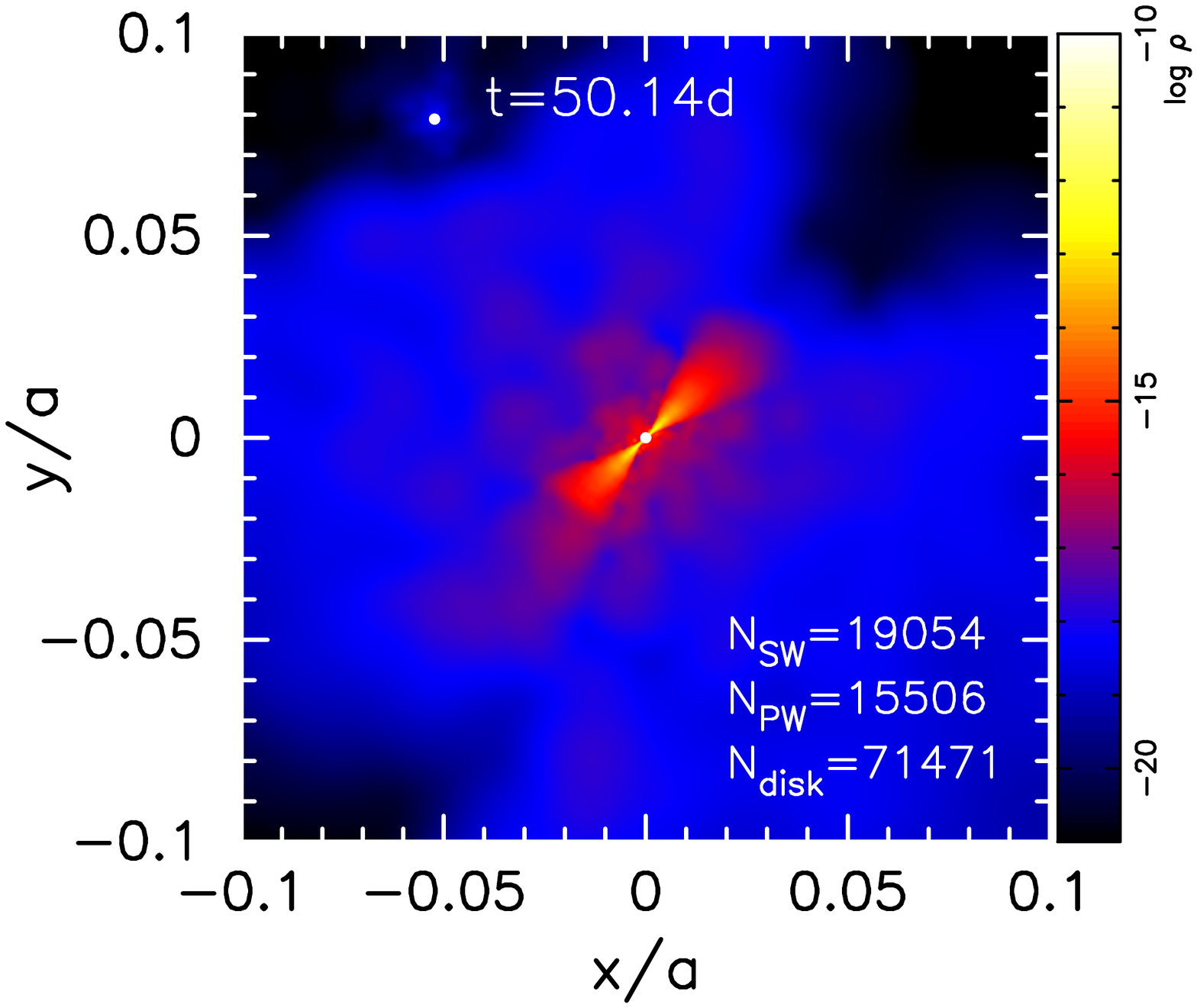}}
    \caption{Snapshots taken from the simulation for Model~1. The upper panels show the column density along the binary orbit axis, whereas the lower panels display the density in the binary
      orbit plane, where we have reversed the directions of the x- and
      y-axes to follow the convention that puts periastron on the
      right hand side.
. They are at $t \sim -50\,\mathrm{d}$, $t \sim -10\,\mathrm{d}$, $t \sim +10\,\mathrm{d}$, and $t \sim +50\,\mathrm{d}$ from left to right, respectively. Annotated at the lower-right part of each panel are, from top to bottom, the numbers of stellar wind particles, pulsar wind particles, and disk particles.}
    \label{fig:snapshots_Model1}
\end{figure*}

\begin{figure*}
	\vspace{0.5cm}
	\resizebox{\hsize}{!}{
	\includegraphics[height=38mm,angle=0]{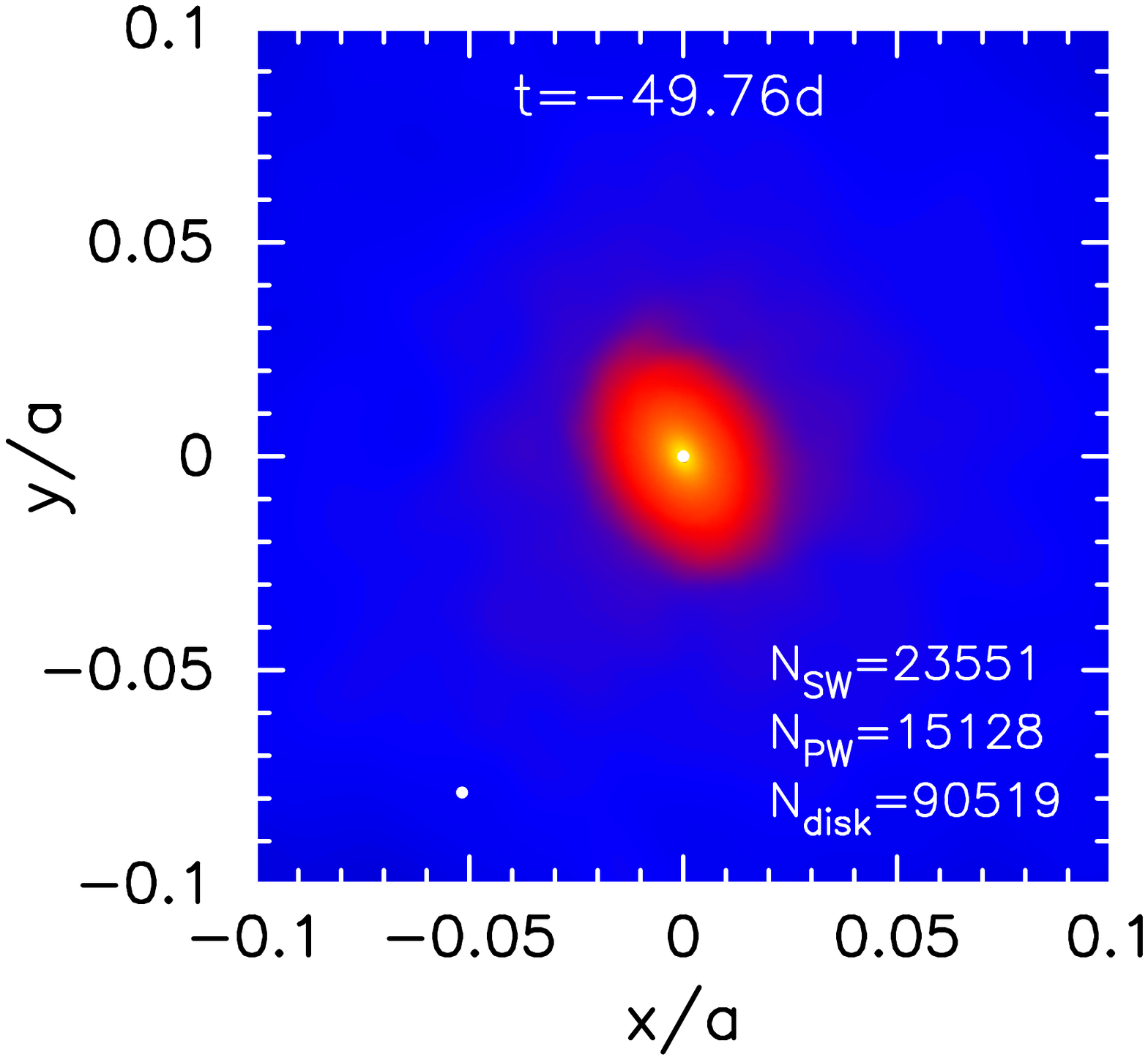} 
	\includegraphics[height=38mm,angle=0]{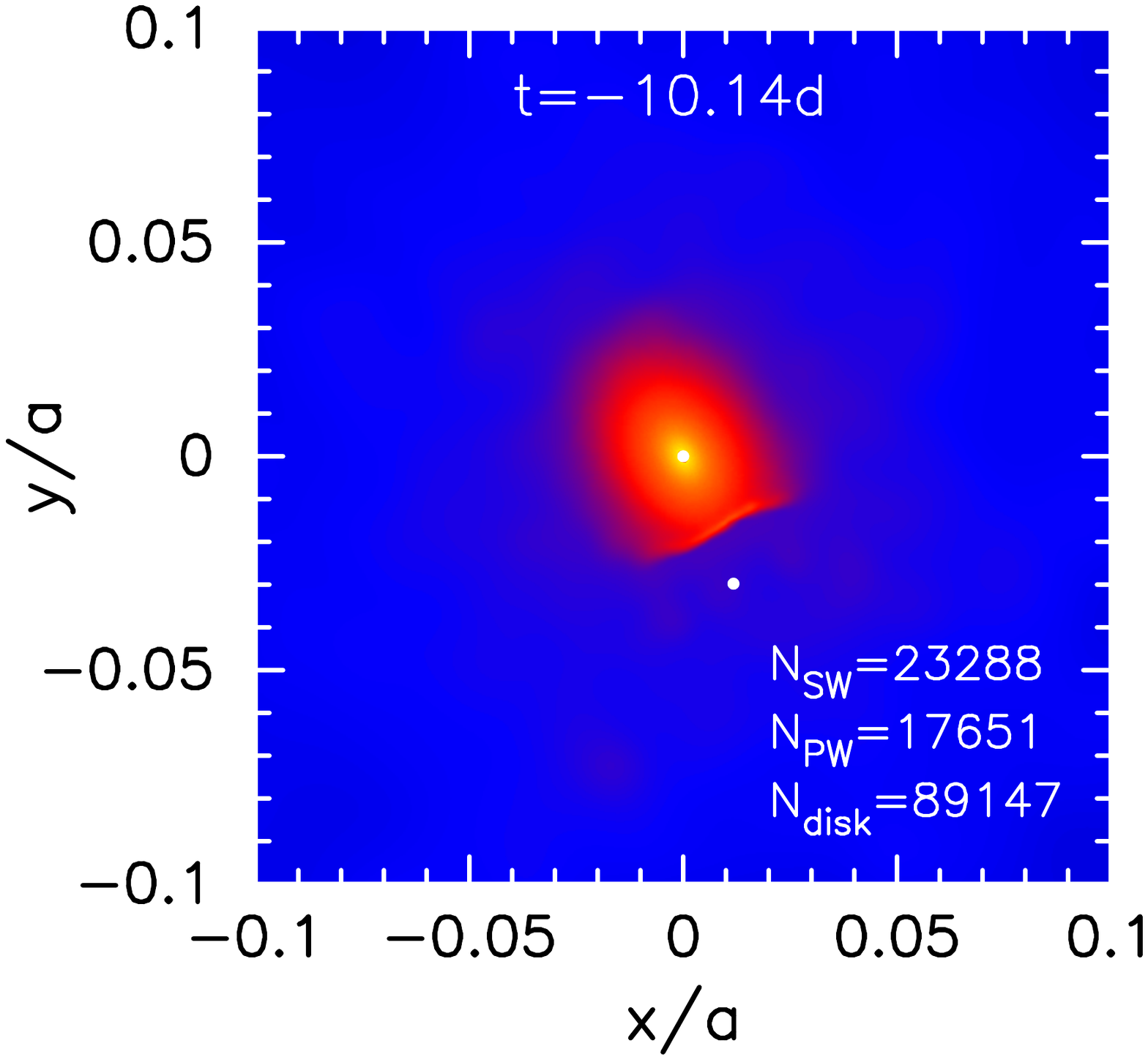} 
	\includegraphics[height=38mm,angle=0]{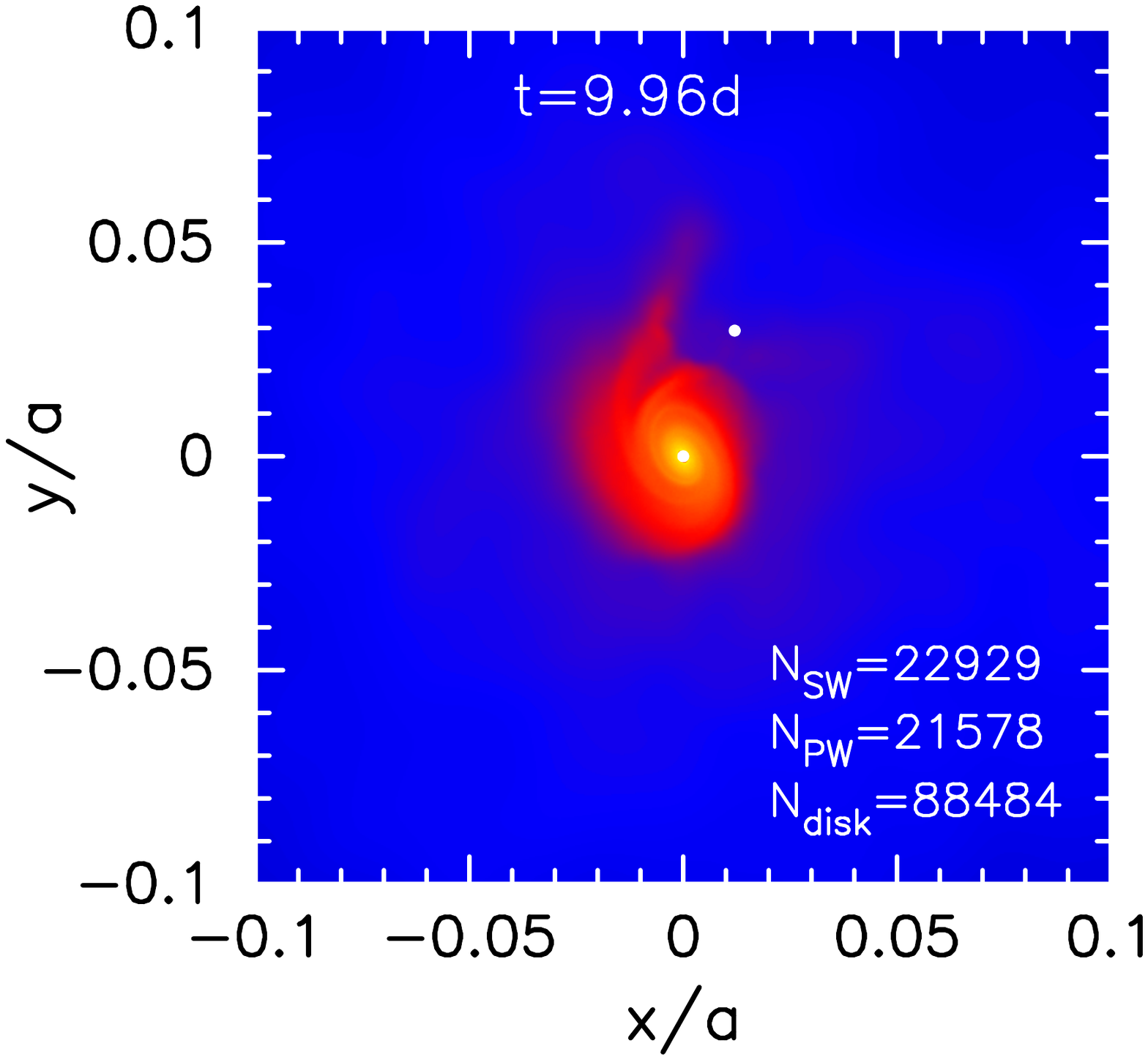} 
	\includegraphics[height=38mm,angle=0]{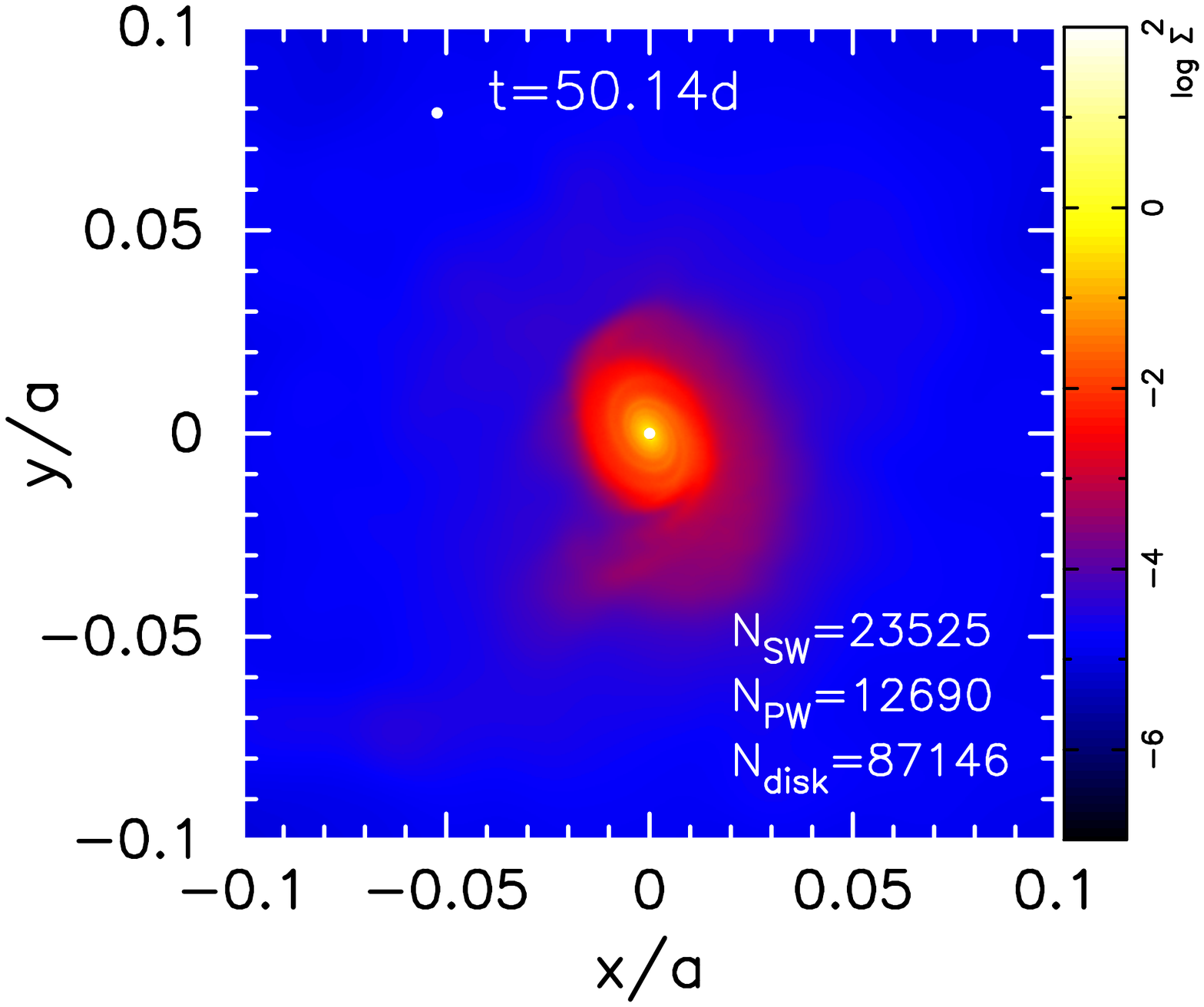}}
	\resizebox{\hsize}{!}{
	\includegraphics[height=38mm,angle=0]{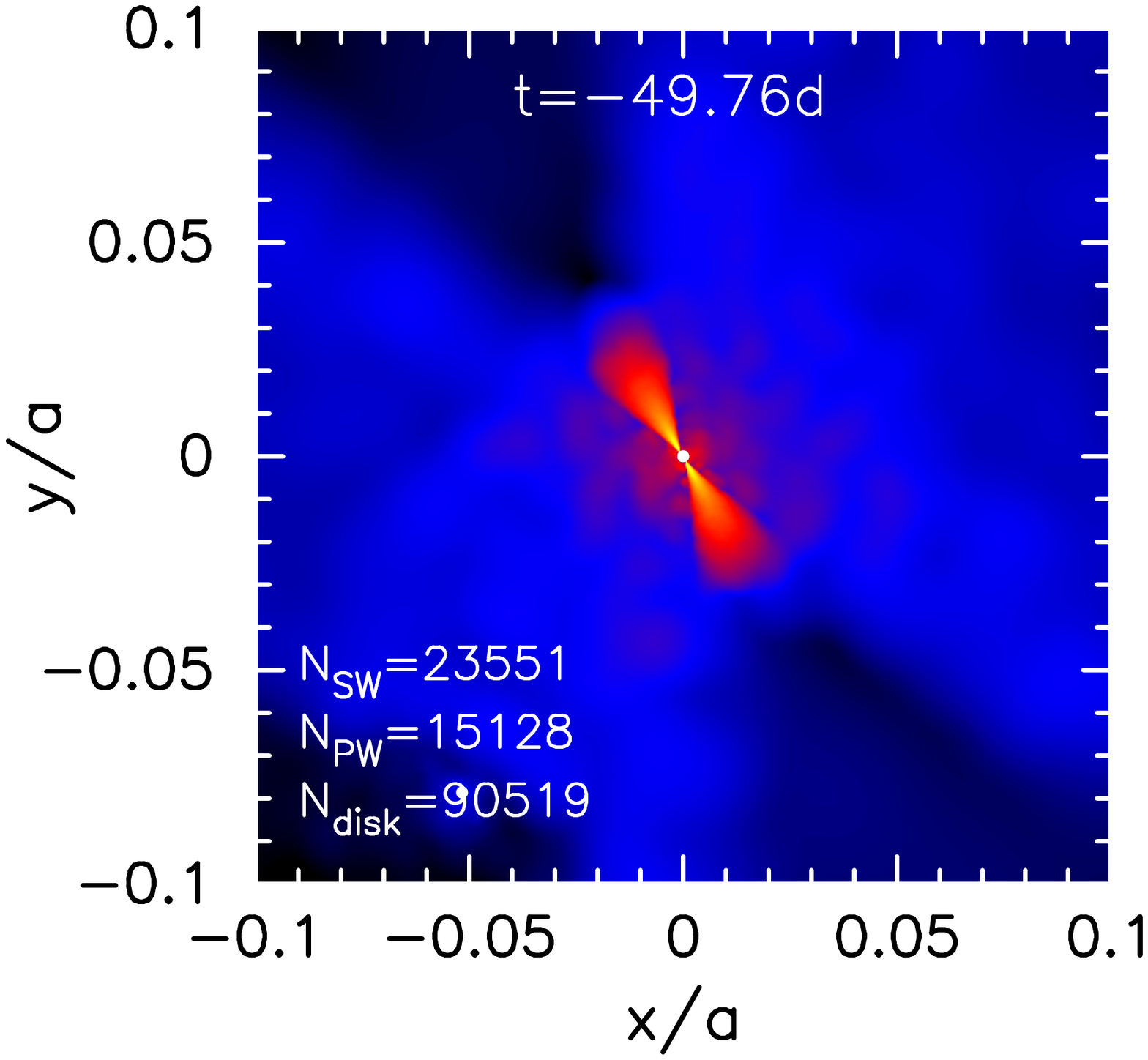} 
	\includegraphics[height=38mm,angle=0]{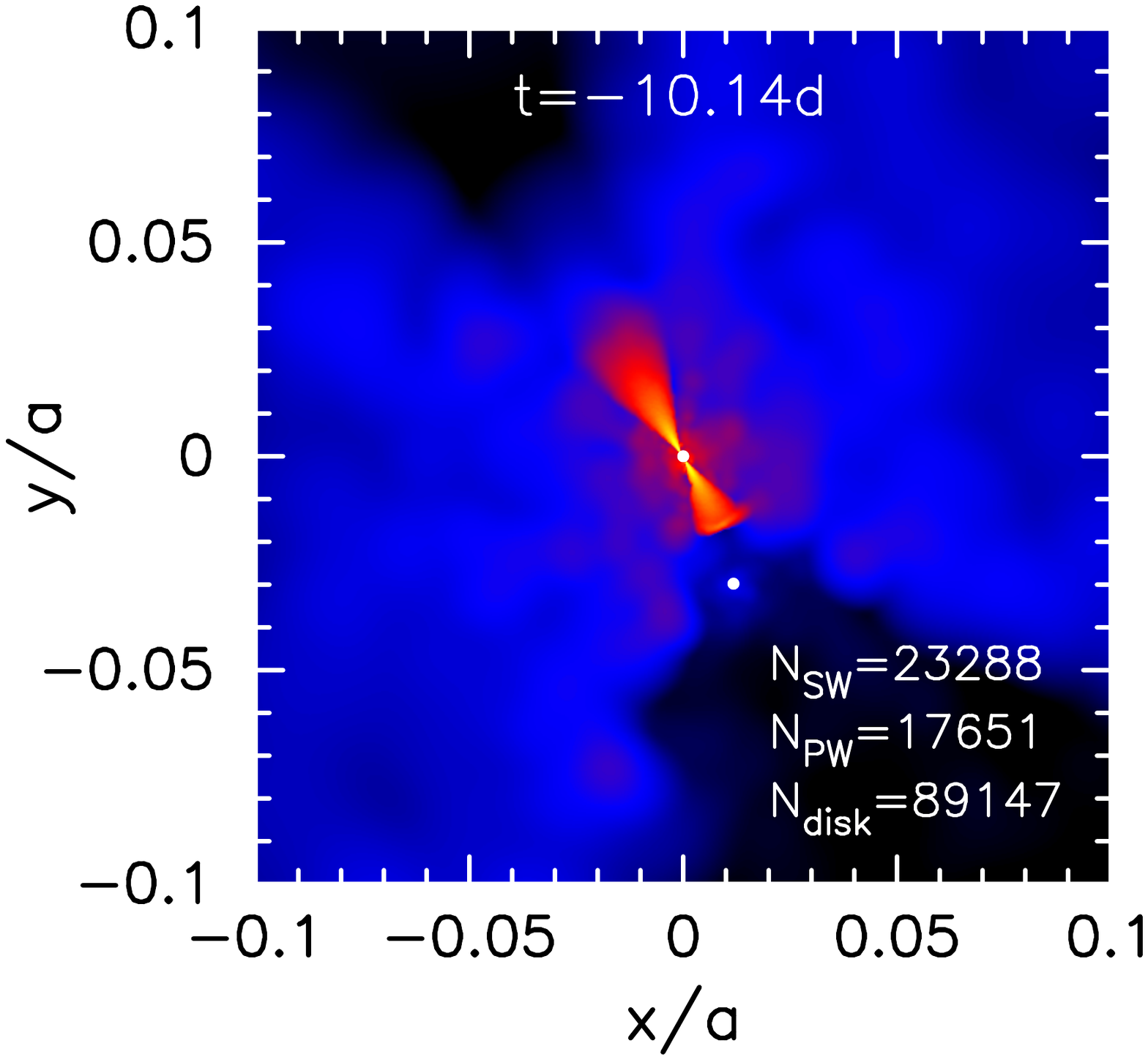} 
	\includegraphics[height=38mm,angle=0]{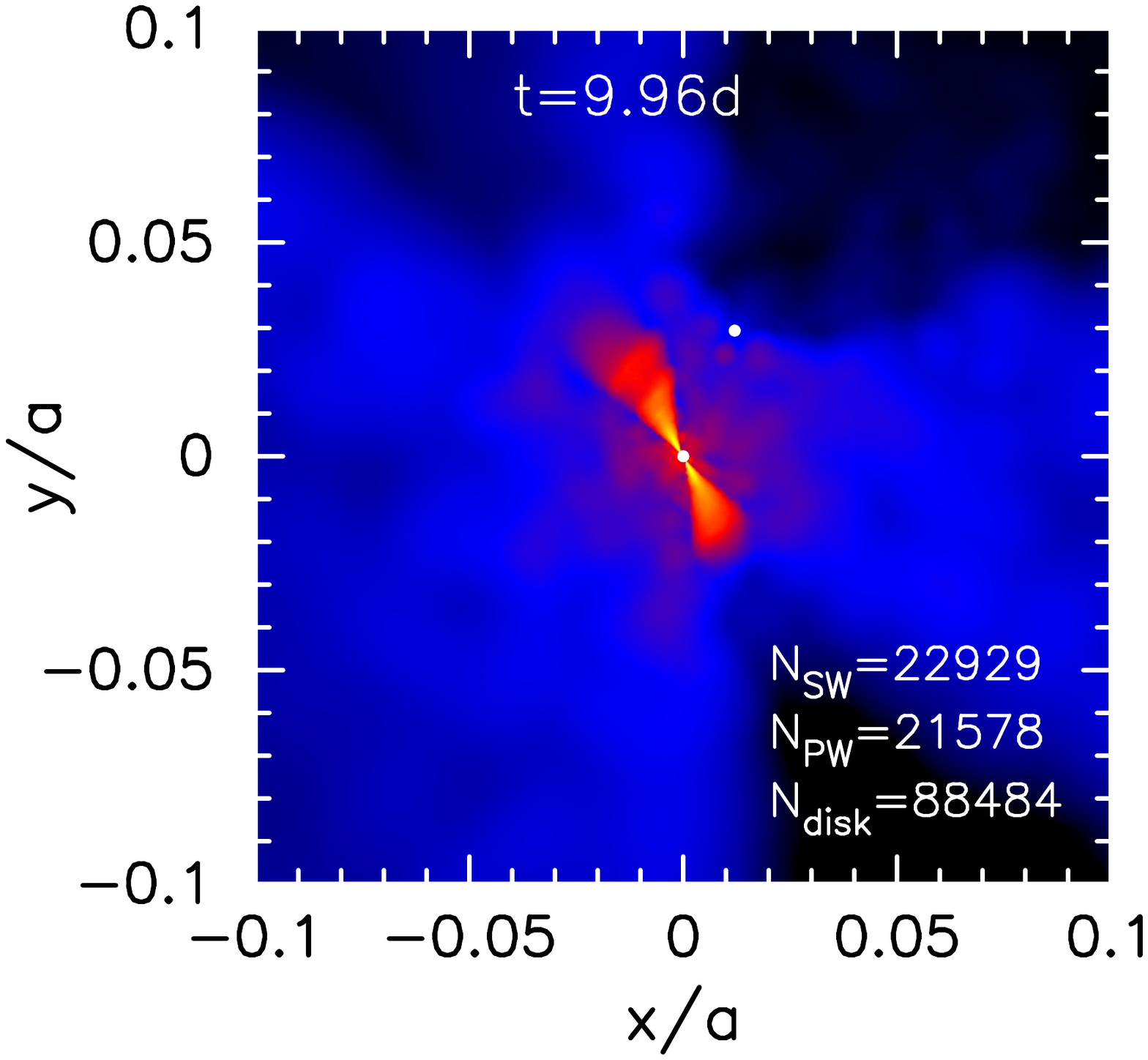} 
	\includegraphics[height=38mm,angle=0]{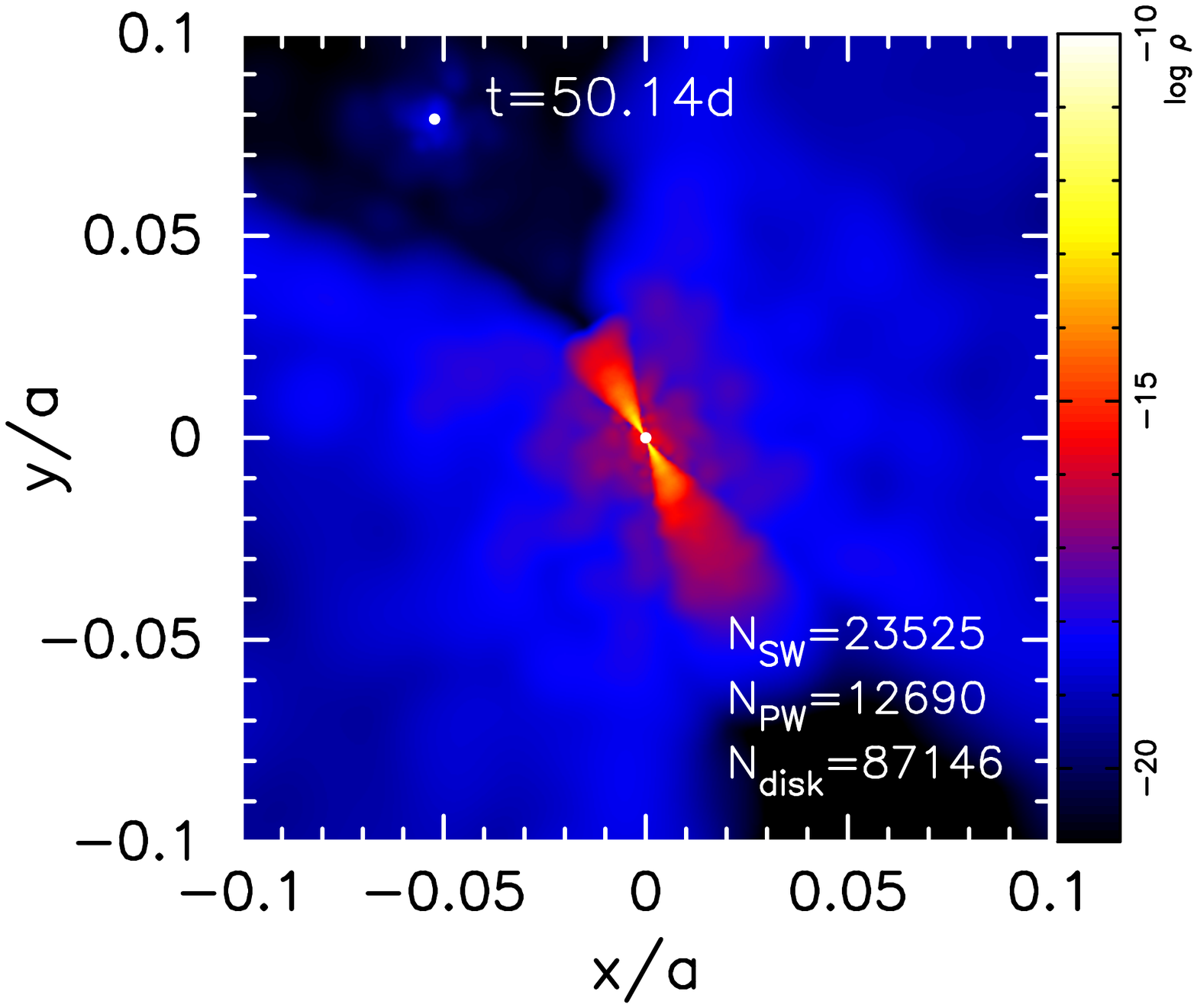}}
    \caption{Same as Fig.~\ref{fig:snapshots_Model1}, but for Model~2.}
    \label{fig:snapshots_Model2}
\end{figure*}

\begin{figure*}
	\vspace{0.5cm}
	\resizebox{\hsize}{!}{
	\includegraphics[height=38mm,angle=0]{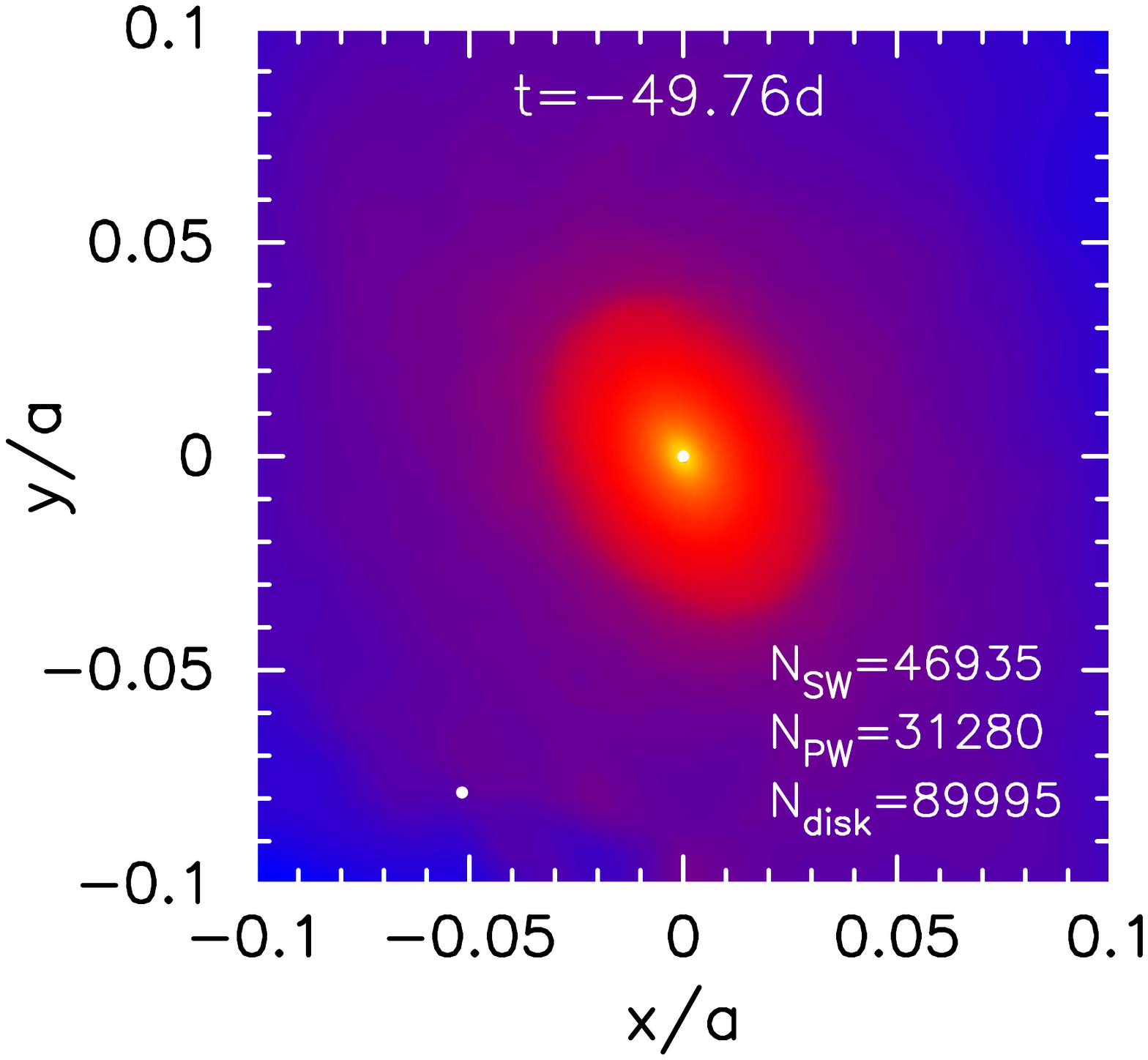} 
	\includegraphics[height=38mm,angle=0]{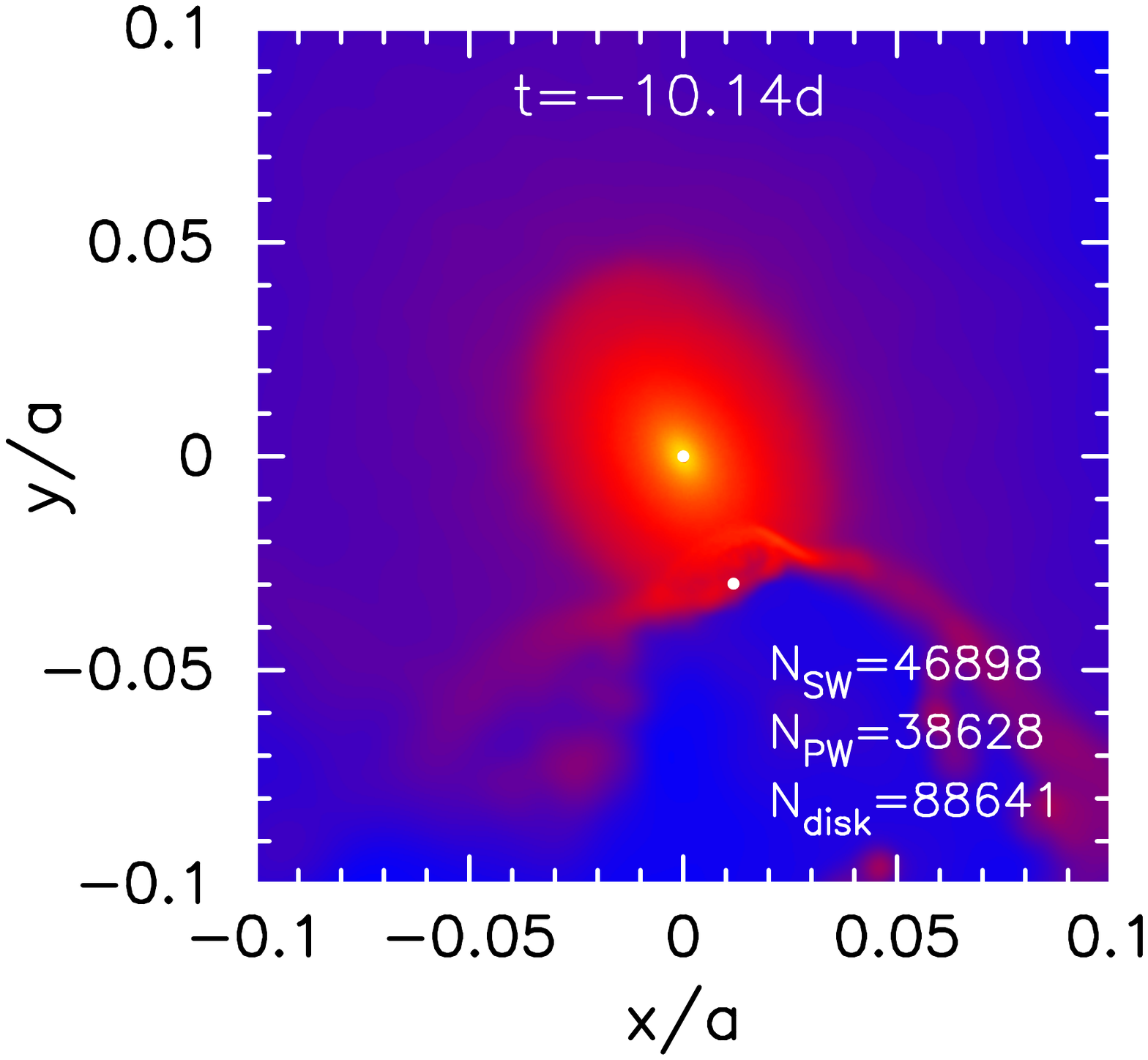} 
	\includegraphics[height=38mm,angle=0]{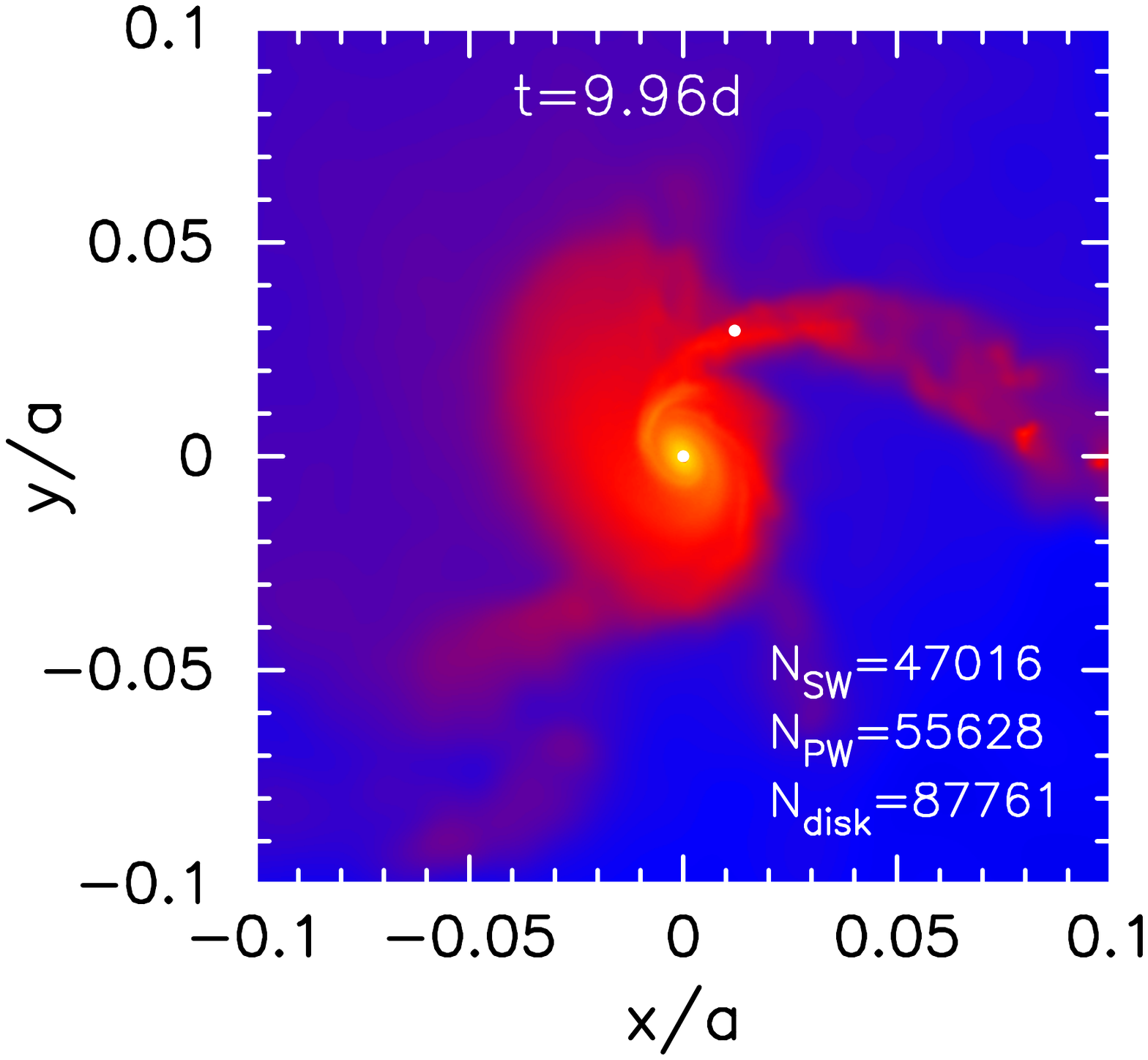} 
	\includegraphics[height=38mm,angle=0]{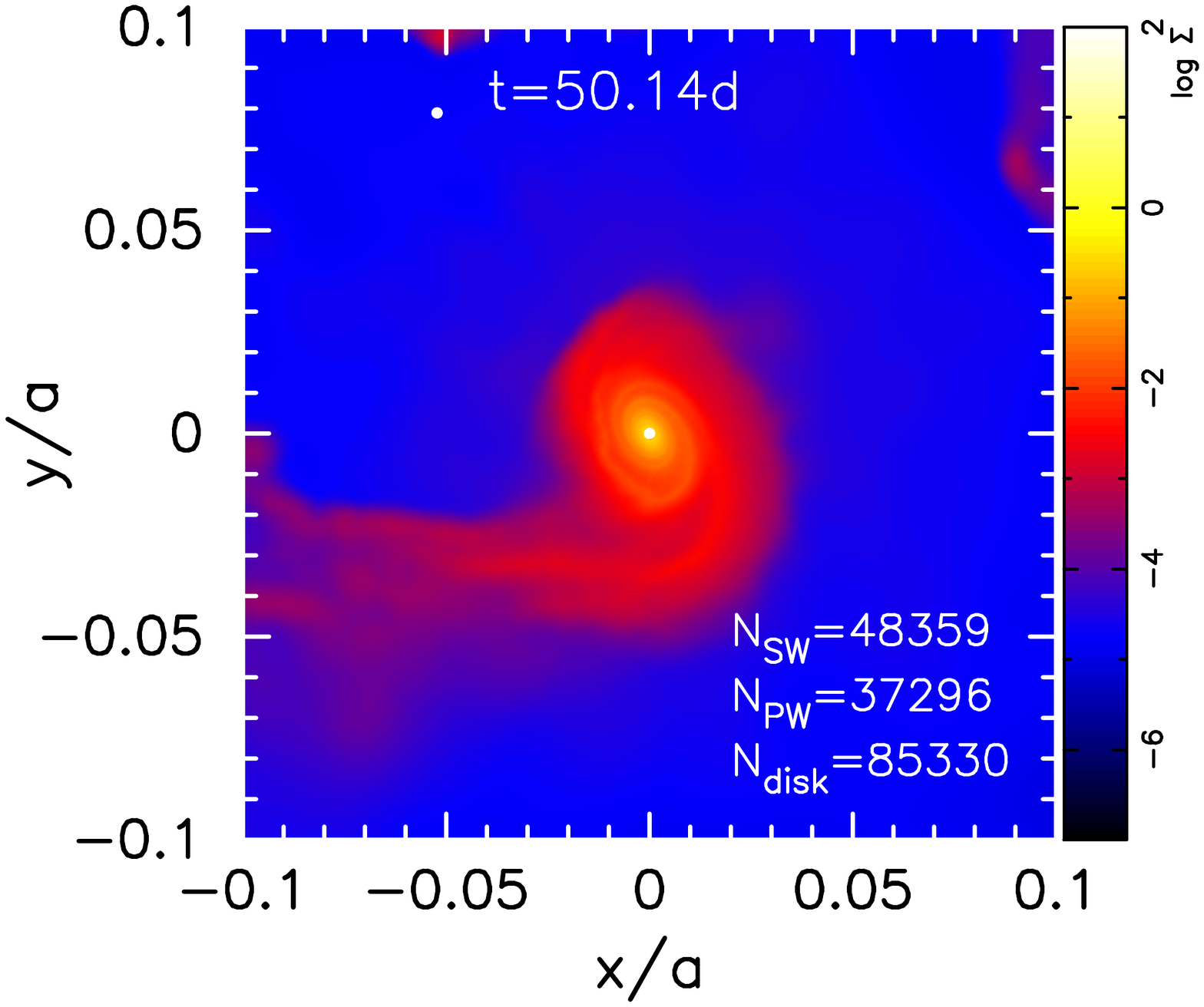}}
	\resizebox{\hsize}{!}{
	\includegraphics[height=38mm,angle=0]{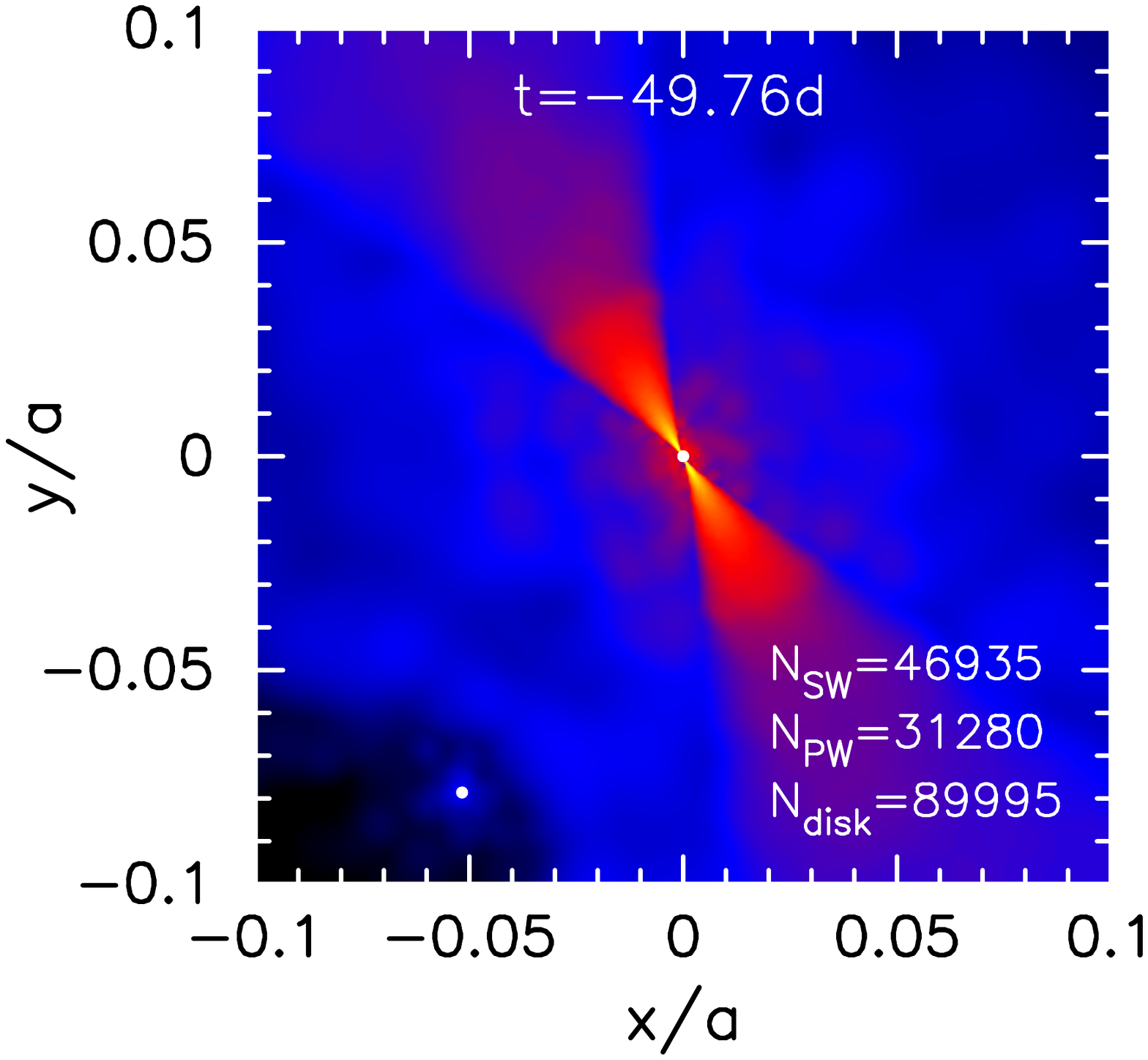} 
	\includegraphics[height=38mm,angle=0]{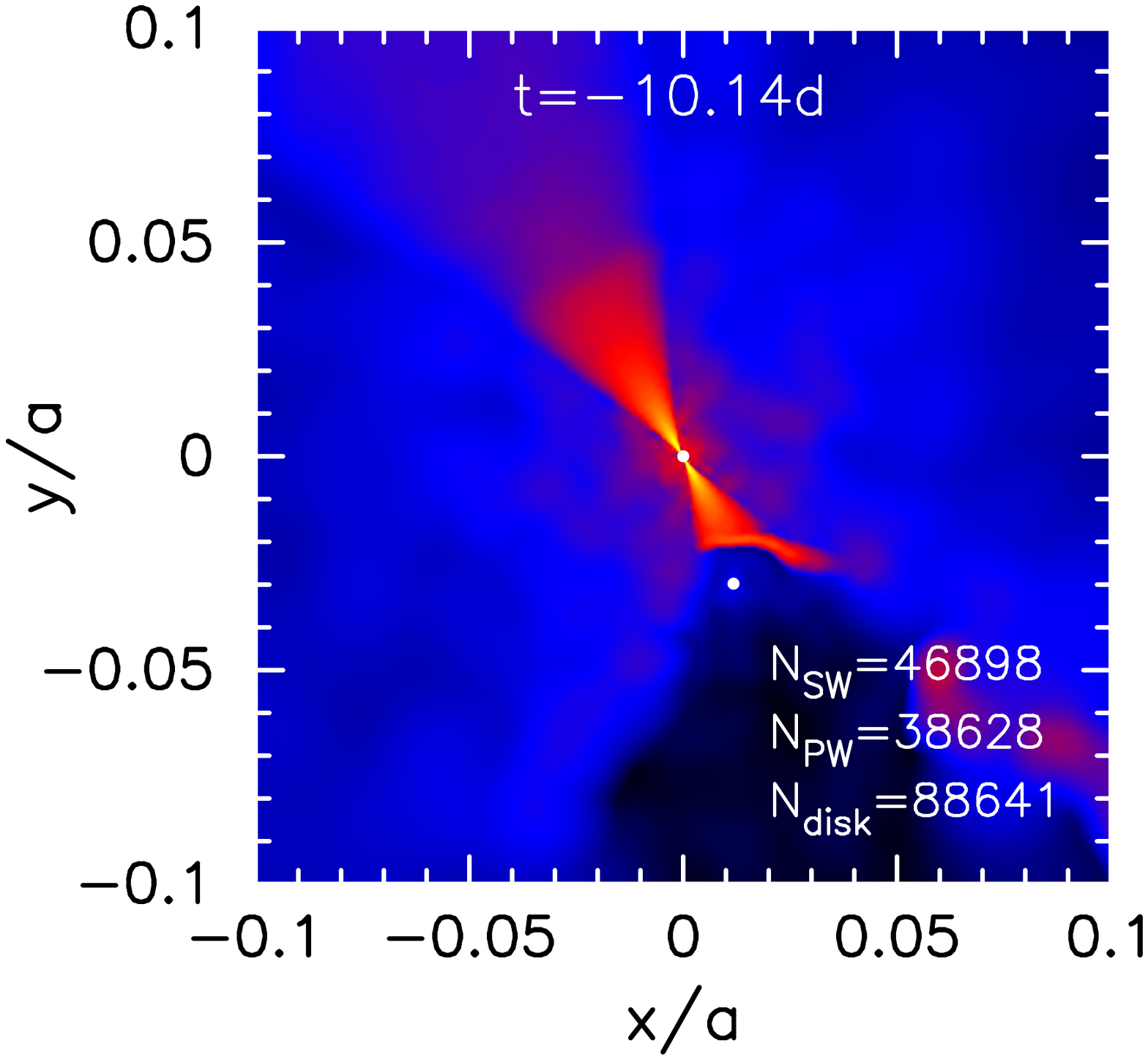} 
	\includegraphics[height=38mm,angle=0]{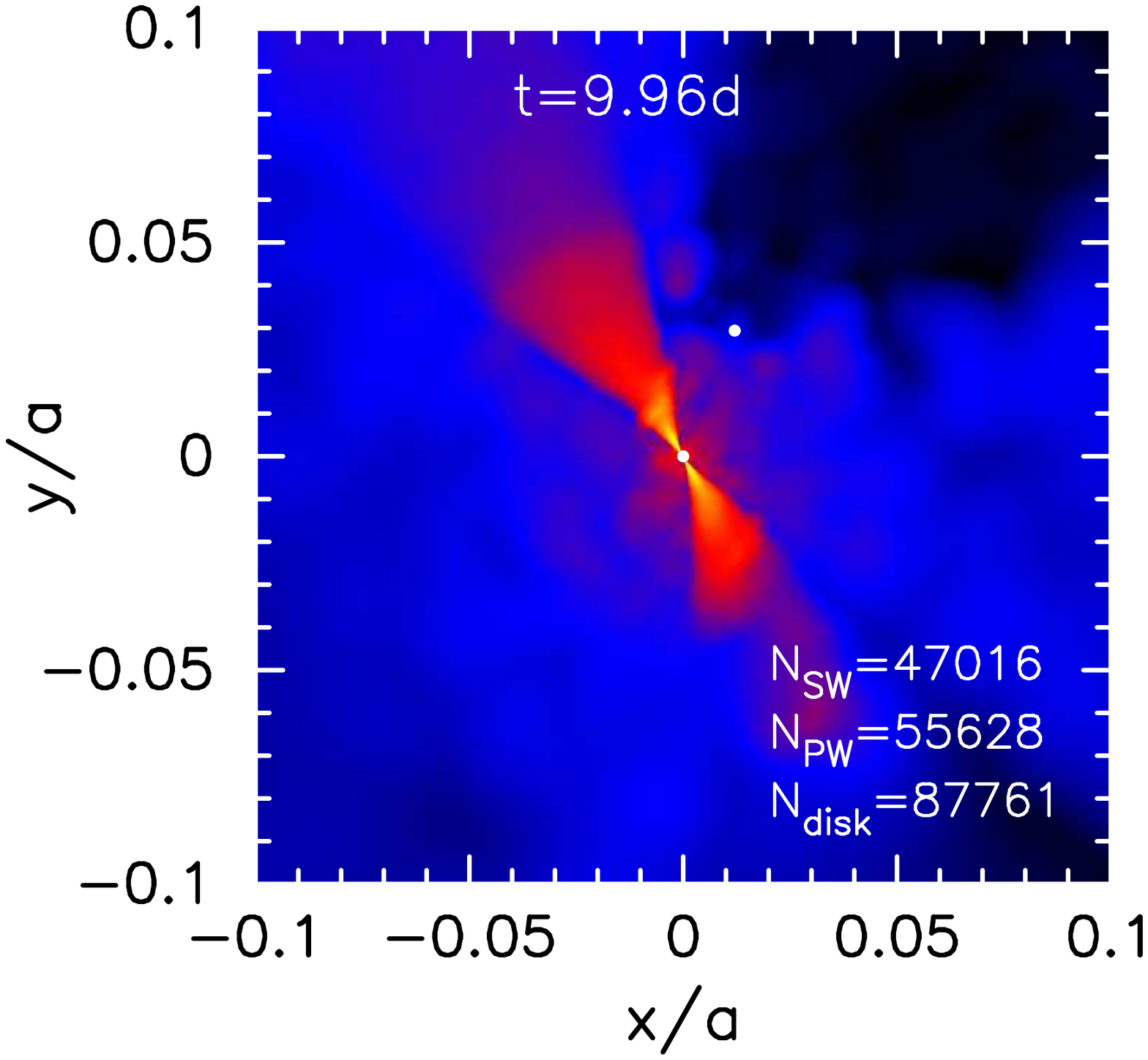} 
	\includegraphics[height=38mm,angle=0]{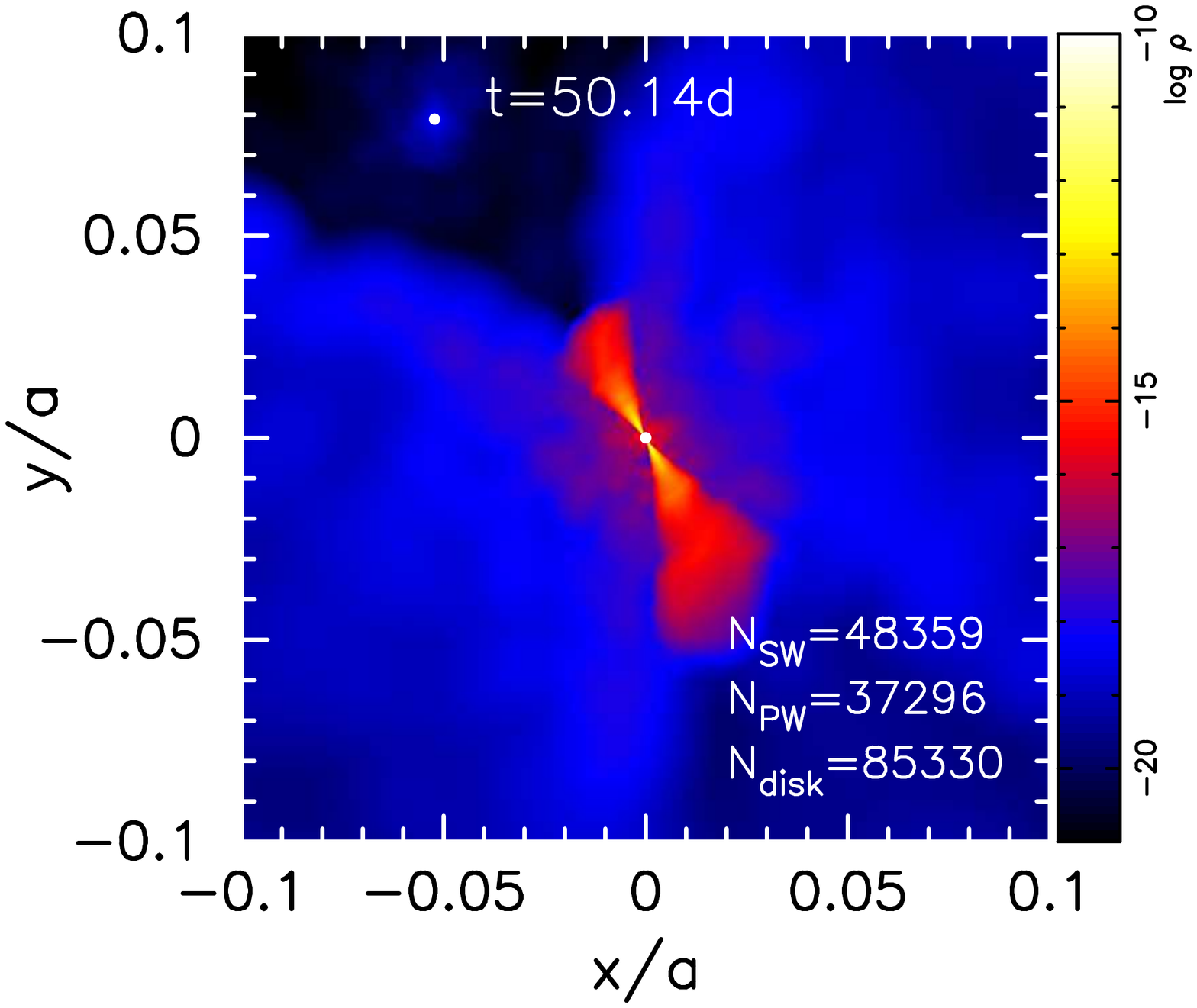}}
    \caption{Same as Fig.~\ref{fig:snapshots_Model1}, but for Model~3.}
    \label{fig:snapshots_Model3}
\end{figure*}

Figures~\ref{fig:snapshots_Model1}--\ref{fig:snapshots_Model3} present several snapshots taken from the simulations for Models~1--3, respectively. In each figure, the upper panels show the column density along the binary orbit axis, whereas the lower panels display the density in the binary orbit plane. The snapshots are taken at $t \sim -50\,\mathrm{d}$, $t \sim -10\,\mathrm{d}$, $t \sim +10\,\mathrm{d}$, and $t \sim +50\,\mathrm{d}$ from left to right, respectively. In each panel, the white circle at the center is the Be star and the small white dot moving around it is the pulsar. The Be disk, the Be-star's wind, and the pulsar wind can be easily distinguished by eye, because of large differences in their densities.

In Model~1 (Fig.~\ref{fig:snapshots_Model1}), where the Be disk has $\beta = 45^\circ$ and $\gamma = -45^\circ$, the pulsar is initially in the region where the wind of the Be star cannot reach owing to the shadowing presence of the Be disk, so that no head-on collision between the pulsar wind and the Be wind occurs (first snapshot). The pulsar stays in the disk shadow until $t \sim -30\,\mathrm{d}$. At this stage, the interaction surface between the pulsar wind and the disk outer radius covers only a small solid angle around the pulsar, implying that only a small fraction of the pulsar wind power is available for particle acceleration during this stage. At $t \sim -30\,\mathrm{d}$, however, the pulsar exits the disk shadow. After this the pulsar wind directly collides with the Be wind, forming an interaction surface covering a much larger solid angle around the pulsar than in the previous disk-shadow stage (second snapshot). This colliding-wind stage lasts until a few days after periastron, when the pulsar enters the shadow of the Be disk again. This second disk-shadow stage lasts only briefly. At $t \sim +10\,\mathrm{d}$ (third snapshot), the pulsar exits the disk shadow, so that the pulsar wind resumes direct collision with the Be wind, which lasts until the end of the simulation (last snapshot).

In contrast, in Model~2 (Fig.~\ref{fig:snapshots_Model2}), the pulsar initially moves in the stellar wind (first snapshot). The ram pressure of the pulsar wind in the wind-wind collision region increases with decreasing distance between the pulsar and the Be star. This lasts until $t \sim -20\,\mathrm{d}$ when the pulsar enters the shadow region of the Be disk. As mentioned above, the interaction surface between the pulsar wind and the Be disk is small in this stage (second snapshot).
Then, a few days before periastron, the pulsar exits the shadow region and resumes direct collision with the Be wind (third snapshot). This strong interaction phase continues until $t \sim +20\,\mathrm{d}$ when the pulsar enters the disk shadow again and stays there for the next $\sim 100\,\mathrm{d}$, during which the interaction between the pulsar wind and the Be disk is very weak (last snapshot).

In Model~3 (Fig.\ref{fig:snapshots_Model3}), the initial disk has the same disk parameters as in Model~2, except that the disk radius is five times larger. The interaction between the pulsar and the Be star in Model~3 is, at first glance, very similar to that in Model~2: The pulsar wind directly collides with the Be wind for $t \lesssim -20\,\mathrm{d}$, $0\,\mathrm{d} \lesssim t \lesssim +20\,\mathrm{d}$, and $t \gtrsim +120\,\mathrm{d}$, whereas the pulsar is in the shadow region of the Be disk for $-20\,\mathrm{d} \lesssim t \lesssim 0\,\mathrm{d}$ and $+20\,\mathrm{d} \lesssim t \lesssim +120\,\mathrm{d}$. A closer inspection, however, exposes a subtle difference. In Model~3, when the pulsar is in the disk shadow, the pulsar wind termination shock, where the non-thermal X-ray emission originates, forms at a position much closer to the pulsar, and so has a larger solid angle around the pulsar, than in Model~2. We will see how this difference in the shock geometry influences the non-thermal X-ray flux in the next section.

\subsection{Emission Model for Shocked Pulsar Wind}
\label{sec:emission-model}

We adopt the same emission model as \citet{Takata2012} to calculate the non-thermal X-ray emission at each orbital phase, using data from the above hydrodynamic simulations. 
First, the simulation volume is divided into $201^{3}$ uniform grids, and at each grid point the pulsar wind pressure is calculated.
Then, from the 3-D distribution of the pulsar wind pressure, synchrotron emission is locally evaluated by using an assumption for the local magnetic fields (for the calculation scheme, see \citep{Takata2012}).
In these calculations, we implicitly assume that particles are accelerated through the first-order Fermi mechanism at the pulsar wind termination shock, and have a power-law energy distribution as $f(\Gamma) \sim \Gamma^{-p}$, where $\Gamma$ is the Lorentz factor and $f(\Gamma)$ is the number density of shocked particles per Lorentz factor at each grid point.
To calculate the energy distribution of the accelerated particles we adopt $p=2$ as the canonical value, and assume that 50\% of the total pulsar wind pressure is due to the magnetic fields. 
The total emission is obtained by integrating the local emissions over the whole simulation volume.

\begin{figure*}
\includegraphics[width= 17cm,angle=0]{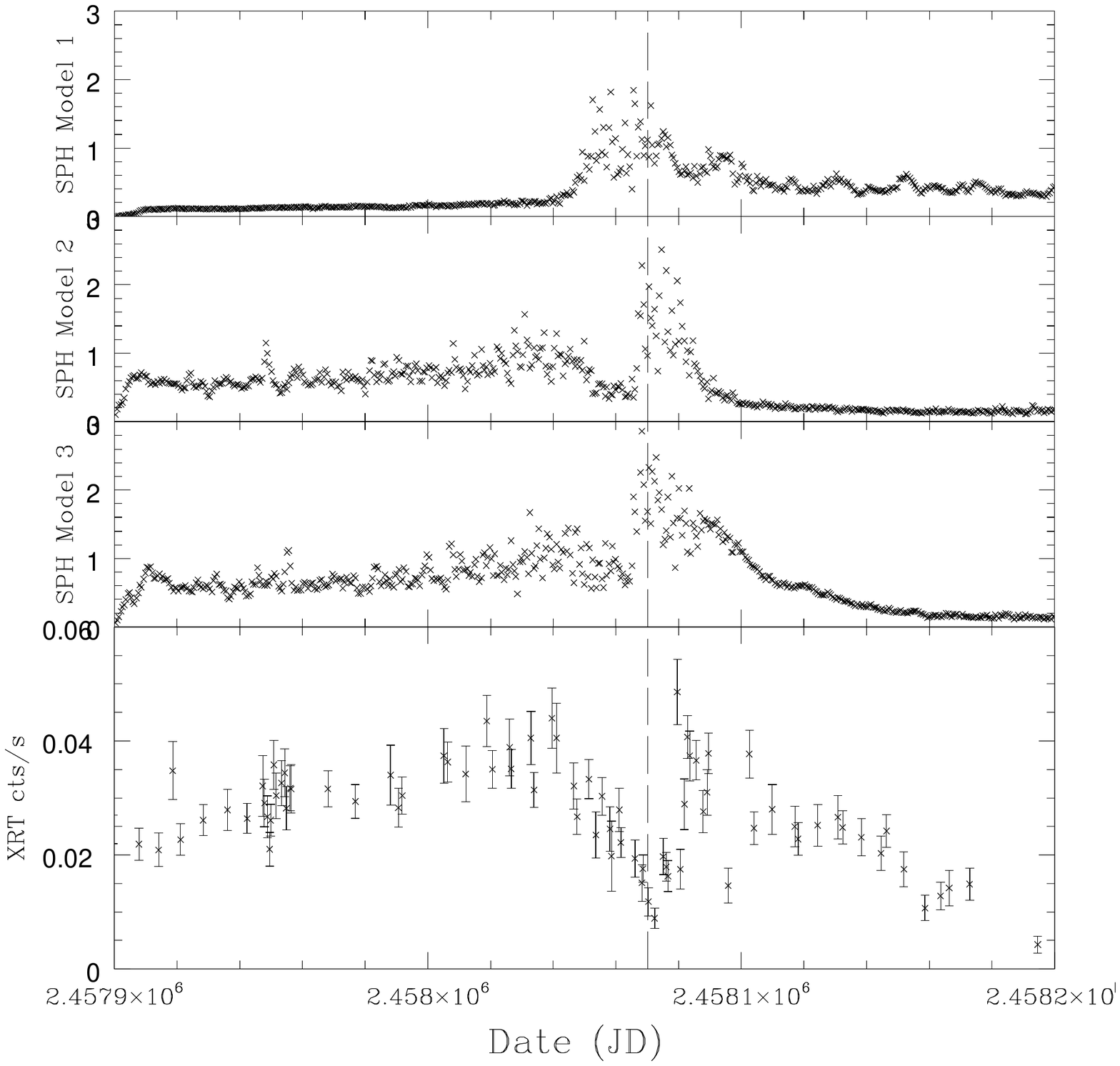} 
    \caption{Simulated X-ray light curves for the three SPH models compared to the Swift XRT count rate (bottom panel). The models are in flux units of $10^{-12}$ erg/sq.cm-s. The date of periastron was JD 2458070 and is indicated by the vertical dashed line. }
    \label{fig:model_xray_lc}
\end{figure*}

The resulting simulated X-ray light curves for three models are compared in Fig.~\ref{fig:model_xray_lc}.
From the difference between the Model~1 (top panel) and Model~2 (middle panel) light curves, we can easily see how the azimuthal direction of the Be circumstellar disk, $\gamma$, and the Be-wind velocity, $V_\mathrm{wind}$, affect the non-thermal X-ray flux.
In these models with typical disk density and size, the X-ray flux is much stronger when the pulsar wind collides with the stellar wind than when it does with the Be disk. 
Note that the wind velocity affects only the former flux, increasing the shocked pulsar wind pressure. 
As a result, in Model~1, the X-ray flux is low until the pulsar exits the disk shadow and starts interacting with the Be wind at $t \sim -30\,\mathrm{d}$. Then, an outburst occurs, of which the X-ray flux has a broad peak around periastron and gradually declines afterwards, with a short break due to the second transit of the pulsar through the disk shadow region. 

In contrast, in Model~2, the continuing collision between the pulsar wind and the Be wind causes gradual increase in the X-ray flux towards the peak at periastron, which is interrupted for $\sim 20\,\mathrm{d}$ just before periatron, when the pulsar moves through the disk shadow. 
After reaching the peak around periastron, the X-ray flux starts to go down rapidly at $t \sim 20\,\mathrm{d}$, when the pulsar enters the disk shadow, to a level much lower than the pre-perastron level. This low state continues until the end of the simulation.
Note that the characteristic features of the Model 2 light
      curve, particularly those prior to periastron, globally agree
      with the observed X-ray variations, except that the model
      predicts the occurrence of the X-ray dip slightly earlier than
      the observation. This discrepancy in the timings of the
      simulated and observed X-ray dip may be resolved by a small
      change in the disk geometry adopted for Model 2, i.e., with the
      azimuthal direction $\gamma$ slightly larger than $30^\circ$ 
      and/or the tilt angle $\beta$ slightly smaller than $45^\circ$.

On the other hand, the effect of the initial disk radius on the X-ray flux is rather subtle. 
At first glance, the X-ray light curves for Model~2 ($r_\mathrm{disk} = 0.02a \sim 16 R_\mathrm{OB}$) and Model~3 ($r_\mathrm{disk} = 0.1a \sim 81 R_\mathrm{OB}$) are very similar: a gradual increase of flux towards the peak around periastron, interrupted by a brief dip just before periastron, and a subsequent, long-lasting low state. 
A closer look, however, reveals that there are differences at epochs when the X-rays mainly originate from the collision of the pulsar wind with the disk outer part. 
In Model~3, the dip feature prior to periastron is less conspicuous than in Model~2, with small flares that partly fill in the dip. 
Moreover, the post-periatron decline is much slower in Model~3 than in Model~2.
These differences arise from the fact that the timescale for the pulsar to cross the disk is comparable to that to strip the disk gas around the orbit.
Since our simulation is non-relativistic, where the pulsar wind has the same momentum flux as, but much smaller inertia than, the relativistic wind does \citep[e.g.,][]{Bosch-Ramon2012}, the differences between the simulated X-ray light curves for Model~2 and Model~3 disk parameters could turn out to be much smaller in relativistic simulations.

In summary, from the comparison between the observed X-ray light curve and those calculated on the basis of SPH simulations, it is found that the gradual increase of flux before $t \sim -30\,\mathrm{d}$ and the peak around periastron are due to the collision between the pulsar wind and the wind of the Be star, whereas the dip just before periastron and the low X-ray state after periastron occur when the pulsar is in the shadow of the Be disk, where no direct wind-wind collision takes place.
The disk orientation and density are also roughly constrained to be $\beta \sim 45^\circ$ and $\gamma \sim 30^\circ$ (see Fig.~\ref{fig:geometry} for the definition of these angles) and $\rho_0 \sim 10^{-11}\,\mathrm{g\;cm}^{-3}$, respectively.
The disk size, however, is not well constrained from our non-relativistic simulations.

\subsection{Free-Free Absorption in the Radio}
\label{sec:radio-absorption}

In \cite{lyne2019}, the monitoring observations of radio pulses at 1520~MHz show that the radio pulsation started weakening just before periastron and then were gone for 2$-$3 weeks from several days after periastron. It is of interest to see whether either model discussed above, particularly Model~2, is consistent with this radio behavior. Below we calculate the free-free absorption at 1520~MHz along the line of sight to the pulsar, using the SPH simulation data.

The free-free absorption coefficient at a radio frequency $\nu$ is given by 
\begin{equation}
\alpha_\mathrm{ff} = \frac{2^{5/2} \pi^{1/2} e^6}{(3mkT)^{3/2} c} 
       \frac{Z^2 n_\mathrm{i} n_\mathrm{e}}{\nu^2} \bar{g}
\label{eq:alpha_ff}
\end{equation}
\citep[e.g.,][]{Rybicki1979}, where $m$ is the electron mass, $Z$ is the atomic charge, $n_\mathrm{i}$ and $n_\mathrm{e}$ are the ion and electron densities, respectively, $T$ is the temperature of the circumstellar gas, and $\bar{g}$ is the average Gaunt factor, which is given by
\begin{equation}
\bar{g} = \frac{3^{1/2}}{\pi} \left[ \ln \frac{(2kT)^{3/2}}{\pi e^2 Z m \nu}
       - \frac{5 \gamma_\mathrm{E}}{2} \right]
\label{eq:gbar}
\end{equation}
for $\nu \gg \nu_\mathrm{pe}$, where $\nu_\mathrm{pe} \sim 9 \times 10^3 n_\mathrm{e}^{1/2}\,\mathrm{Hz}$ is the plasma frequency and $\gamma_\mathrm{E} \simeq 0.5772$ is Euler's constant \citep{Spitzer1978}.
Then, the optical depth, $\tau_\mathrm{ff}$, is calculated by numerically integrating $\alpha_\mathrm{ff}$, using the simulation data, along the line of sight to the pulsar as
\begin{equation}
    \tau_\mathrm{ff} = \int \alpha_\mathrm{ff} \, ds,
    \label{eq:tau_nu}
\end{equation}
with $ds$ being the line element along the line of sight. 
We adopt that the longtitude of periastron is $40^\circ$ \citep{ho2017}. The inclination angle of the binary orbit is assumed to be $70^\circ$.

\begin{figure}
	\resizebox{\hsize}{!}{\includegraphics{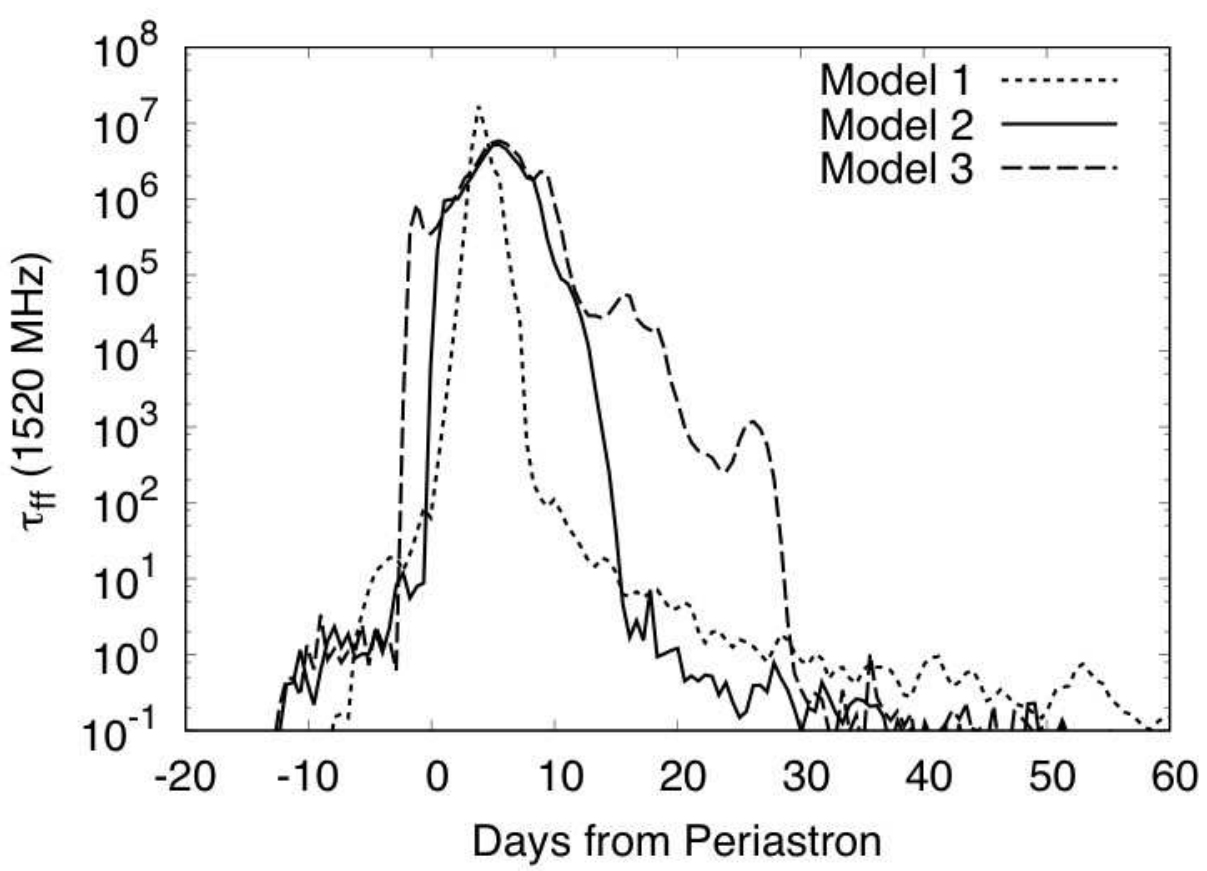}}
    \caption{Optical depth for the free-free absorption at 1520~MHz along the line of sight to the pulsar. The dotted, the solid, and the dashed lines are for Models 1, 2, and 3, respectively.}
    \label{fig:tau_radio}
\end{figure}

Figure~\ref{fig:tau_radio} shows the variations of the optical depth for the free-free absorption at 1520~MHz for Models~1-3, from which a few interesting features are observed. First, in all models the optical depth becomes very large for a while after periastron. This large optical depth is due to the occultation of the pulsar by the Be disk. Specifically, the condition of $\tau_\mathrm{ff} \gg 1$ necessary for completely obscuring pulsed emission is met from $t \sim -5\,\mathrm{d}$ to $t \sim +20\,\mathrm{d}$ in Model~1 (dotted line), from $t \sim -3\,\mathrm{d}$ to $t \sim +15\,\mathrm{d}$ in Model~2 (solid line), and from $t \sim -3\,\mathrm{d}$ to $t \sim +28\,\mathrm{d}$ in Model~3 (dashed line). Second, the transition between the optically thin and thick absorption states occurs more gradually in Model~1 than in Models~2 and 3, owing to the difference in disk orientation. Finally, the duration of the optically thick absorption state is longer for a larger disk size. Particularly, Model~3 predicts too-long optically thick state to be consistent with the radio monitoring observation.

In conclusion, Model~2, of which the simulated X-ray light curve is in best agreement with the X-ray observations, also provides the variation pattern of the free-free absorption that is most consistent with the radio observations.

\section{Discussion}

\subsection{Pulsar wind contributions}

\begin{figure}
	\hspace{-0.3cm}
	\includegraphics[width=80mm,angle=0]{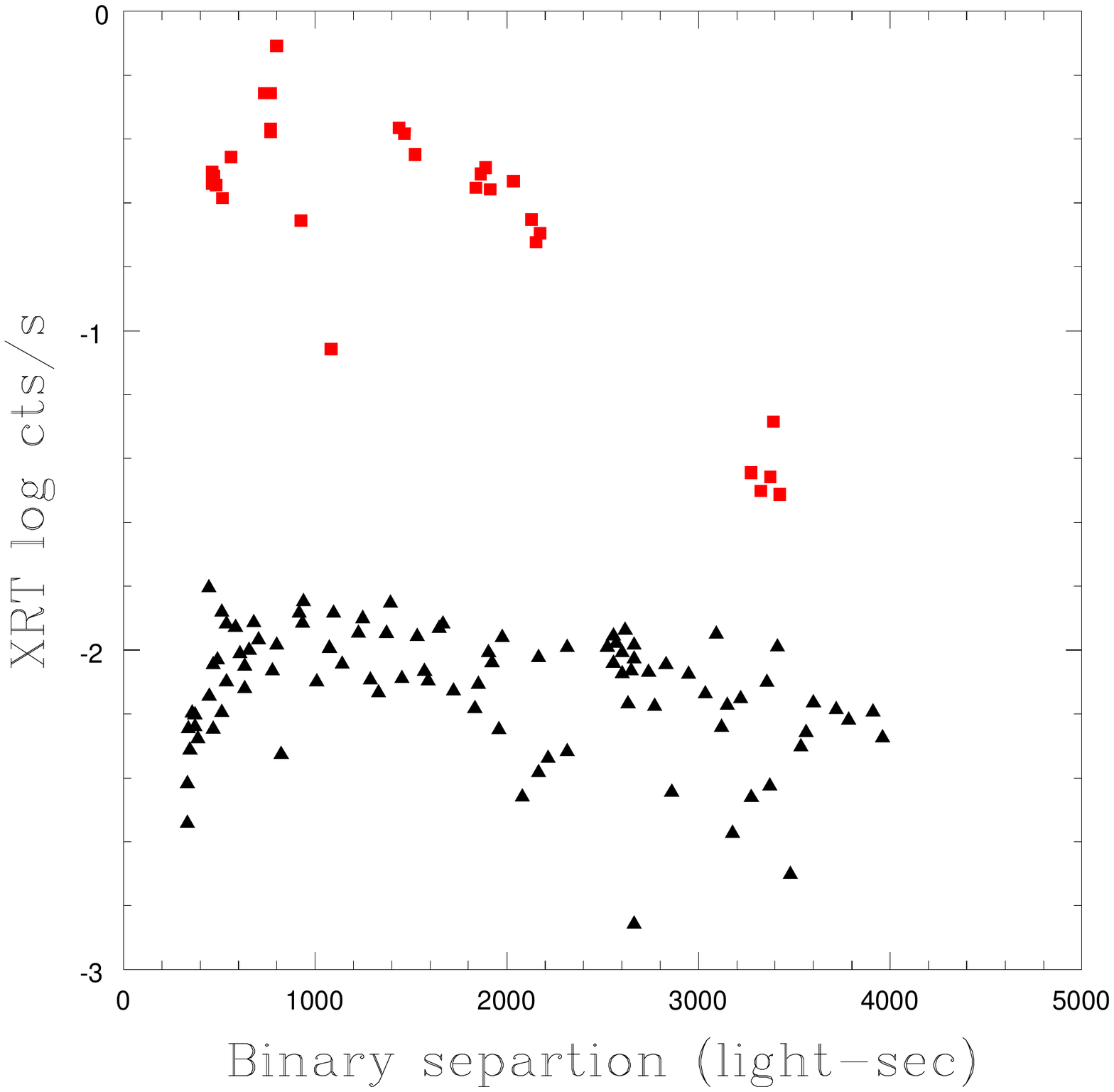}
    \caption{\textit{Neil Gehrels Swift Observatory} X-ray flux versus binary separation. The black triangles show measurements for PSR J2032+4127 reduced by a factor of 3.1 to compensate for the relative distances between this system and PSR B1259-63 (red squares).}
    \label{fig:compare}
\end{figure}

We can compare the long-term X-ray flux measured by \textit{Neil Gehrels Swift Observatory} for the system presented in this paper with the only other known similar system, PSR B1259-63 during a periastron passage. PSR J2032+4127 is at a distance of 1.3 kpc, whereas PSR B1259-63 lies 2.3 kpc away. So if we want to directly compare the X-ray luminosities we need to divide all the values for PSR J2032+4127 by a factor of 3.1 to correct for the relative distances. \textit{Neil Gehrels Swift Observatory} measurements for PSR B1259-63 were extracted from the \textit{Neil Gehrels Swift Observatory} archive for the periastron passage that took place in Dec 2010. The binary separation values for the partners in PSR B1259-63 were obtained using the published orbital elements and assuming an orbital inclination of 23 degrees and stellar mass ratio of 14.2 \citep{Shannon2014}. The result of the scaled comparison is shown in Figure~\ref{fig:compare}, from which it can be immediately seen that PSR B1259-62 is significantly more X-ray bright than PSR J2032+4127.

The luminosity from a pulsar wind is believed to stem from the loss of rotational energy from a classic dipole field which is given by Equation \ref{eq:rot}:

\begin{equation}
\label{eq:rot}
\frac{dE}{dt}=\frac{2}{3c^2}{M_{B}^{2}}{\Omega^{4}}{sin^{2}}\alpha
\end{equation}
where $M_{B}$ is the magnetic dipole moment, $\Omega$ is the spin frequency and $\alpha$ is the angle between the rotation and magnetic axis. Assuming that $M_{B}$ and $\alpha$ are the same for both systems, this implies that the energy in the pulsar wind should simply scale by the factor of $\Omega^{4}$. So putting in spin periods of 48ms for PSR B1259-63 and 143ms for PSR J2032+4127 implies a flux ratio of $\sim$76. Observationally the scaled luminosity difference between the two systems is a factor of $\sim$40. So assuming that the magnetic dipole moment does not vary much between neutron stars the explanation for this factor of $\sim$2 difference may lie with the parameter $\alpha$.

\subsection{Coronal X-rays}

Coronal X-ray emission is expected from isolated OB stars in the range $10^{30} - 10^{32}$ erg/s \citep{cassinelli1994, naze2011}. So there may well be such a contribution in addition to any energy from the colliding winds. The peak observed Swift count rate of $\sim$0.04 cts/s corresponds to a 2 - 10 keV X-ray luminosity of $\sim 4 \times 10^{32}$ erg/s assuming $N_H = 7.7 \times 10^{-21} cm^{-2}$ cm and $\Gamma = 2.0$ (values from \cite{ho2017}) and a distance of 1.38 kpc. The baseline upper limit on the luminosity, well before periastron, was estimated assuming an XRT count rate limit of $\le$0.004 cts/s. This corresponds to an limit on the luminosity of $\sim 5 \times 10^{31}$ erg/s. The latter value would be consistent with coronal X-ray emission from a star of this type, B0Ve, so the increased X-ray luminosity due to the interaction between the two stars in the system peaks at the order of $\sim 3.5 \times 10^{32}$ erg/s, 

Incidentally the optical object in PSR B1259-63 is very similar with a classification of O9.5Ve, so similar coronal X-ray emission would be expected.

\section{Conclusions}

We have presented detailed optical spectroscopy and X-ray photometry from the first observed periastron passage of the semicentennial binary system PSR J2032+4127. This rare event has allowed us to study the detailed interaction of a neutron star and a circumstellar disk plus stellar wind around an OB star in a setting where normally this doesn't occur. Our detailed SPH models have allowed us to have a better understanding of the probable orbital configuration of the system and gain a better insight into the behaviour of this colliding wind system.

\section*{Acknowledgements}

We are grateful to the {\it Neil Gehrels Swift Observatory} for the valuable X-ray coverage of this system. The Liverpool Telescope is operated on the island of La Palma by Liverpool John Moores University in the Spanish Observatorio del Roque de los Muchachos of the Instituto de Astrofisica de Canarias with financial support from the UK Science and Technology Facilities Council.
We are also grateful to Jumpei Takata for giving us his code for calculating synchrotron emission using SPH data, and 
we thank Tsuguya Naito for allowing us to adapt his original figure to create  Fig.~\ref{fig:geometry} in this work. Fermi work at NRL is supported by NASA.

\bsp

\label{lastpage}

\begin{thebibliography}{99}

\bibitem[\protect\citeauthoryear{Abdo et al.}{2009}]{abdo2009} Abdo A.~A., et al., 2009, Sci, 325, 848

\bibitem[\protect\citeauthoryear{Abeysekara et al.}{2018}]{abey2018} Abeysekara A.~U., et al., 2018, ApJ, 867, L19 

\bibitem[\protect\citeauthoryear{Aharonian et al.}{2002}]{aha2002} Aharonian F., et al., 2002, A\&A, 393, L37 
\bibitem[\protect\citeauthoryear{Aharonian et al.}{2005}]{aha2005} Aharonian F., et al., 2005, A\&A, 431, 197

\bibitem[\protect\citeauthoryear{Aliu et al.}{2014}]{aliu2014} Aliu E., et al., 2014, ApJ, 783, 16 

\bibitem[\protect\citeauthoryear{Barnsley, Smith, \& Steele}{2012}]{frodo-pipe} Barnsley R.~M., Smith R.~J., Steele I.~A., 2012, AN, 333, 101

\bibitem[\protect\citeauthoryear{{Bate}, {Bonnell}  \& {Price}}{{Bate} et~al.}{1995}]{Bate1995}
{Bate} M.~R.,  {Bonnell} I.~A.,   {Price} N.~M.,  1995, \mnras, {277, 362}

\bibitem[\protect\citeauthoryear{Bednarek, Banasi{\'n}ski, \& Sitarek}{2018}]{bednarek2018} Bednarek W., Banasi{\'n}ski P., Sitarek J., 2018, JPhG, 45, 015201

\bibitem[\protect\citeauthoryear{{Bosch-Ramon}, {Barkov}, {Khangulyan}  \&
  {Perucho}}{{Bosch-Ramon} et~al.}{2012}]{Bosch-Ramon2012}
{Bosch-Ramon} V.,  {Barkov} M.~V.,  {Khangulyan} D.,   {Perucho} M.,  2012,
  \aap, 544, A59
  
  \bibitem[\protect\citeauthoryear{Cassinelli et al.}{1994}]{cassinelli1994} Cassinelli J.~P., Cohen D.~H., Macfarlane J.~J., Sanders W.~T., Welsh B.~Y., 1994, ApJ, 421, 705

\bibitem[\protect\citeauthoryear{Camilo et al.}{2009}]{camilo} Camilo F., et al., 2009, ApJ, 705, 1 

\bibitem[\protect\citeauthoryear{{Carciofi} \& {Bjorkman}}{{Carciofi} \& {Bjorkman}}{2006}]{Carciofi2006}{Carciofi} A.~C.,  {Bjorkman} J.~E.,  2006, \apj, {639, 1081}

\bibitem[\protect\citeauthoryear{Dubus}{2013}]{dubus2013} Dubus G., 2013, A\&ARv, 21, 64 

\bibitem[\protect\citeauthoryear{Dubus \& Cerutti}{2013}]{2013A&A...557A.127D} Dubus G., Cerutti B., 2013, A\&A, 557, A127 

\bibitem[\protect\citeauthoryear{Cassinelli et al.}{1994}]{1994ApJ...421..705C} Cassinelli J.~P., Cohen D.~H., Macfarlane J.~J., Sanders W.~T., Welsh B.~Y., 1994, ApJ, 421, 705 

\bibitem[\protect\citeauthoryear{{Ferland}}{{Ferland}}{1996}]{Ferland1996}{Ferland} G.~J.,  1996, {Hazy, A Brief Introduction to Cloudy 90}


\bibitem[\protect\citeauthoryear{Gehrels et al.}{2004}]{gehrels2004} Gehrels N., et al., 2004, ApJ, 611, 1005

\bibitem[\protect\citeauthoryear{Hanuschik}{1989}]{han89} Hanuschik R.~W., 1989, Ap\&SS, 161, 61 

\bibitem[\protect\citeauthoryear{Hanuschik}{1994}]{han94} Hanuschik R.~W., 1994, Ap\&SS, 216, 99 https://www.overleaf.com/project/

\bibitem[\protect\citeauthoryear{Ho et al.}{2017}]{ho2017} Ho W.~C.~G., Ng C.-Y., Lyne A.~G., Stappers B.~W., Coe M.~J., Halpern J.~P., Johnson T.~J., Steele I.~A., 2017, MNRAS, 464, 1211 

\bibitem[\protect\citeauthoryear{Jennings et al.}{2018}]{gaia} Jennings R.~J., Kaplan D.~L., Chatterjee S., Cordes J.~M., Deller A.~T., 2018, ApJ, 864, 26


\bibitem[Huang(1972)]{huang72} Huang, S.-S.\ 1972, \apj, 171, 549

\bibitem[\protect\citeauthoryear{Jaschek \& Jaschek}{1993}]{jaschek} Jaschek C., Jaschek M., 1993, A\&AS, 97, 807 

\bibitem[\protect\citeauthoryear{Kolka et al.}{2017}]{kolka2017} Kolka I., Eenm{\"a}e T., Laur J., Aret A., 2017, RNAAS, 1, 37 

\bibitem[\protect\citeauthoryear{Li et al.}{2017}]{li2017} Li K.~L., Kong A.~K.~H., Tam P.~H.~T., Hou X., Takata J., Hui C.~Y., 2017, ApJ, 843, 85 

\bibitem[\protect\citeauthoryear{Li et al.}{2018}]{li2018} Li K.~L., Takata J., Ng C.~W., Kong A.~K.~H., Tam P.~H.~T., Hui C.~Y., Cheng K.~S., 2018, ApJ, 857, 123 


\bibitem[\protect\citeauthoryear{Lyne et al.}{2015}]{lyne2015} Lyne A.~G., Stappers B.~W., Keith M.~J., Ray P.~S., Kerr M., Camilo F., Johnson T.~J., 2015, MNRAS, 451, 581 



\bibitem[\protect\citeauthoryear{Massey \& Thompson}{1991}]{mt91} Massey P., Thompson A.~B., 1991, AJ, 101, 1408

\bibitem[\protect\citeauthoryear{Morales-Rueda et al.}{2004}]{frodospec} Morales-Rueda L., Carter D., Steele I.~A., Charles P.~A., Worswick S., 2004, AN, 325, 215 



\bibitem[\protect\citeauthoryear{Naz{\'e} et al.}{2011}]{naze2011} Naz{\'e} Y., et al., 2011, ApJS, 194, 7 

\bibitem[\protect\citeauthoryear{Ng et al.} {2019}]{ng2019} Ng C-Y. et al, 2019, in preparation.


\bibitem[\protect\citeauthoryear{Okazaki}{1997}]{o97} Okazaki A.~T., 1997, A\&A, 318, 548

\bibitem[\protect\citeauthoryear{{Okazaki}, {Bate}, {Ogilvie}  \& {Pringle}}{{Okazaki} et~al.}{2002}]{Okazaki2002}{Okazaki} A.~T.,  {Bate} M.~R.,  {Ogilvie} G.~I.,   {Pringle} J.~E.,  2002, \mnras, {337, 967}

\bibitem[\protect\citeauthoryear{{Okazaki}, {Owocki}, {Russell}  \& {Corcoran}}{{Okazaki} et~al.}{2008}]{Okazaki2008}{Okazaki} A.~T.,  {Owocki} S.~P.,  {Russell} C.~M.~P.,   {Corcoran} M.~F.,
  2008, \mnras, {388, L39}

\bibitem[\protect\citeauthoryear{{Okazaki}, {Nagataki}, {Naito}, {Kawachi}, {Hayasaki}, {Owocki}  \& {Takata}}{{Okazaki} et~al.}{2011}]{Okazaki2011}{Okazaki} A.~T.,  {Nagataki} S.,  {Naito} T.,  {Kawachi} A.,  {Hayasaki} K., {Owocki} S.~P.,   {Takata} J.,  2011, \pasj, {63, 893}

\bibitem[\protect\citeauthoryear{Petropoulou et al.}{2018}]{petro2018} Petropoulou M., Vasilopoulos G., Christie I.~M., Giannios D., Coe M.~J., 2018, MNRAS, 474, L22

\bibitem[\protect\citeauthoryear{Reig}{2011}]{reig} Reig P., 2011, Ap\&SS, 332, 1 

\bibitem[\protect\citeauthoryear{Rybicki \& Lightman}{1979}]{Rybicki1979}
Rybicki G.~B., Lightman A.~P., 1979, Radiative Processes in Astrophysics. New York, Wiley-Interscience

\bibitem[\protect\citeauthoryear{Shannon, Johnston, \& Manchester}{2014}]{Shannon2014} Shannon R.~M., Johnston S., Manchester R.~N., 2014, MNRAS, 437, 3255 

\bibitem[\protect\citeauthoryear{Spitzer}{1978}]{Spitzer1978}
Spitzer L., 1978, Physical Processes in the Interstellar Medium. New York, Wiley-Interscience

\bibitem[\protect\citeauthoryear{Stappers et al.} {2019}]{lyne2019} Stappers B. et al., 2019 in preparation.

\bibitem[\protect\citeauthoryear{Steele, Negueruela, \& Clark}{1999}]{be-stars} Steele I.~A., Negueruela I., Clark J.~S., 1999, A\&AS, 137, 147

\bibitem[\protect\citeauthoryear{Steele et al.}{2004}]{lt} Steele I.~A., et al., 2004, Proc. SPIE, 5489, 679 

\bibitem[\protect\citeauthoryear{{Takata} et~al.,}{{Takata} et~al.}{2012}]{Takata2012}{Takata} J.,  et~al., 2012, \apj, {750, 70}

\bibitem[\protect\citeauthoryear{Takata et al.}{2017}]{2017ApJ...836..241T} Takata J., Tam P.~H.~T., Ng C.~W., Li K.~L., Kong A.~K.~H., Hui C.~Y., Cheng K.~S., 2017, ApJ, 836, 241

\bibitem[\protect\citeauthoryear{Walborn \& Fitzpatrick}{1990}]{walborn} Walborn N.~R., Fitzpatrick E.~L., 1990, PASP, 102, 379

\end{thebibliography}
\end{document}